\documentclass[journal = acsmacro,layout=twocolumn]{achemso}
\usepackage[version=3]{mhchem}
\usepackage{graphicx}
\usepackage{dcolumn}
\usepackage{bm}
\usepackage[usenames, dvipsnames]{color}
\usepackage[utf8]{inputenc}
\usepackage[T1]{fontenc}
\usepackage[english]{babel}
\usepackage{amsmath}
\usepackage{amssymb}
\usepackage{multicol}
\usepackage{multirow} 
\usepackage{array}
\usepackage{tikz}
\usepackage{soul}
\usepackage{ulem}
\usepackage{adjustbox}

\usepackage{xr}
\definecolor{purpleIL}{rgb}{0.71372549, 0.71372549, 0.84705882}
\definecolor{purpleTHF}{rgb}{0.48402922, 0.45433295, 0.71066513}
\definecolor{purpleMelt}{rgb}{0.31164937, 0.11995386, 0.5455594}
\definecolor{lightgreen}{rgb}{0.8, 0.9, 0.4}
\definecolor{green}{rgb}{0.2980392156862745, 0.6490196078431373, 0.4}

\definecolor{n1}{rgb}{0, 0, 0.5}
\definecolor{n2}{rgb}{0.0, 0.69215686, 1}
\definecolor{n4}{rgb}{1, 0.77051561, 0}
\definecolor{n10}{rgb}{0.9456328, 0.02977487, 0}

\author{Tiago Outerelo Corvo}
\affiliation{Université Paris Saclay, Laboratoire Léon Brillouin, UMR 12 CNRS-CEA, CEA-Saclay, 91191, Gif-sur-Yvette, France}
\alsoaffiliation[LPS]{Université Paris-Saclay, CNRS, Laboratoire de Physique des Solides, 91405, Orsay, France}

\author{Antoine Jourdain}
\affiliation{Univ Lyon, Universit\'{e} Lyon 1, CNRS, Ing\'{e}nierie des Mat\'{e}riaux Polymères, UMR 5223, F-69003 Lyon, France}
\author{Shona O'Brien}
\affiliation{Univ Lyon, Universit\'{e} Lyon 1, CNRS, Ing\'{e}nierie des Mat\'{e}riaux Polymères, UMR 5223, F-69003 Lyon, France}
\author{Fr\'{e}d\'{e}ric Restagno}
\affiliation{Université Paris-Saclay, CNRS, Laboratoire de Physique des Solides, 91405, Orsay, France}
\author{Eric Drockenmuller}
\email{eric.drockenmuller@univ-lyon1.fr}
\affiliation{Univ Lyon, Universit\'{e} Lyon 1, CNRS, Ing\'{e}nierie des Mat\'{e}riaux Polymères, UMR 5223, F-69003 Lyon, France}
\author{Alexis Chennevière}
\email{alexis.chenneviere@cea.fr}
\affiliation{Université Paris Saclay, Laboratoire Léon Brillouin, UMR 12 CNRS-CEA, CEA-Saclay, 91191, Gif-sur-Yvette, France}

\title{Multiscale structure of poly(ionic liquid)s in bulk and solutions by small angle neutron scattering}

\keywords{SANS, Polymerized Ionic Liquids}

\begin{document}
\begin{abstract}
	Poly(ionic liquid)s (PILs), similar to their ionic liquid (IL) analogues, present a nanostructure arising from local interactions. The influence of this nanostructure on the macromolecular conformation of polymer chains is investigated for the first time by means of an extensive use of small angle neutron scattering on a series of poly(1-vinyl-3-alkylimidazolium)s  with varying alkyl side-chain length and counter-anion both in bulk and in dilute solutions. Radii of gyration are found to increase with the side-chain length in solution as a consequence of crowding interactions between neighboring monomers. In bulk, however, a non monotonic evolution of the radius of gyration reflects a change in chain flexibility and a potential screening of electrostatic interactions. Additionally at smaller scale, SANS provides an experimental estimation of both the chain diameter and the correlation length between neighboring chains, comparison of which unveils clear evidence of interdigitation of the alkyl side-chains. These structural features bring precious insights in the understanding of the dynamic properties of PILs.
\end{abstract}
\section{Introduction}
Poly(ionic liquid)s (PILs) are a special class of polyelectrolytes bearing ionic liquids (ILs) groups in their repeating units. ILs themselves have been an object of interest of many works in the last decades \cite{triolo_nanoscale_2007, canongia_lopes_nanostructural_2006, russina_morphology_2009, russina2012mesoscopic, rocha2013alkylimidazolium, hardacre_small_2010, annapureddy2010origin, araque2015modern, agebbie_long_2017}. They are salts with a melting point typically below 100 \textdegree C, confered by asymmetric and bulky ion pairs. This grants them a large panel of thermal, physical and electrochemical properties as well as an enhanced ionic conductivity.
The asymmetric cation usually features an amphiphilic structure with polar and apolar moieties leading to a local solvophobic-induced phase separation \cite{russina2012mesoscopic}. The versatility of the IL bulk structure is mainly controlled by the size of the apolar moiety, usually an alkyl chain, and spans from a globular structure in which apolar parts are segregated into small domains to a bicontinuous, sponge-like structure for longer alkyl chains \cite{canongia_lopes_nanostructural_2006, triolo_nanoscale_2007}.

Their polymerized form, PILs, sparked increasing interest as an emerging interdisciplinary topic among polymer chemistry and physics, materials science, catalysis, separation, analytical chemistry, and electrochemistry. Initially, they appeared to play a purely complementary role towards the amplification of the functions of ILs, delivering performances that could not readily be afforded by molecular ILs \cite{mecerreyes_polymeric_2011, mudraboyina_123-triazolium-based_2014, shaplov_recent_2015}. 

Rapid advances in the chemistry and physics of PILs have paved the way to the development of novel and versatile polymer electrolytes that are highly relevant for both applied and fundamental research \cite{yuan_polyionic_2013, yuan_polyionic_2011, eftekhari_synthesis_2017}.  The combination of the variety of available ion pairs and different macromolecular architectures such as linear, branched or cross-linked polymer materials results in a wide range of synthetic routes to develop new materials for various applications including dye-sensitized solar cell, fuel cells, batteries, or sensors\cite{zhao_solvent-free_2011, chen_bis-imidazolium_2012, obadia_poly123-triazoliums:_2016,kim_high-performance_2011, weber_effect_2011, yuan_self-assembly_2011}.
The addition of local interactions inherited from ILs to macromolecules results in a complex and rich panel of chemical and physical properties opening new opportunities to design polymer materials with targeted functionalities which are highly related to both structural and dynamic properties of PILs.
MD simulations and Wide Angle X Ray scattering have shown that such interactions can lead to a local structuration of PILs somehow similar to the one observed in the corresponding ILs by forming ionic clusters \cite{liu_alkyl_2017, liu_direct_2016, delhorbe_unveiling_2017, canongia_lopes_nanostructural_2006}. Such nanostructuration was shown to be independent of the PILs polymerization degree ($N$) above a certain threshold \cite{wieland_structure_2019} and was found to play a significant role in both their ionic conductivity \cite{cruz_correlating_2012, iacob_polymerized_2017, doughty_structural_2019} and rheological behavior \cite{nakamura_viscoelastic_2011, nakamura_dielectric_2012, matsumoto_rheological_2021, zhao_effect_2021}. However, there is, up to now, no evidence of the influence of such local structuration on the macromolecular conformation of PIL chains in bulk and solutions. Describing the overall structure of these peculiar polyelectrolytes could give important insights into the wide physico-chemical properties of PILs, their dynamics and their design.

\begin{scheme}
	\includegraphics[width=\linewidth]{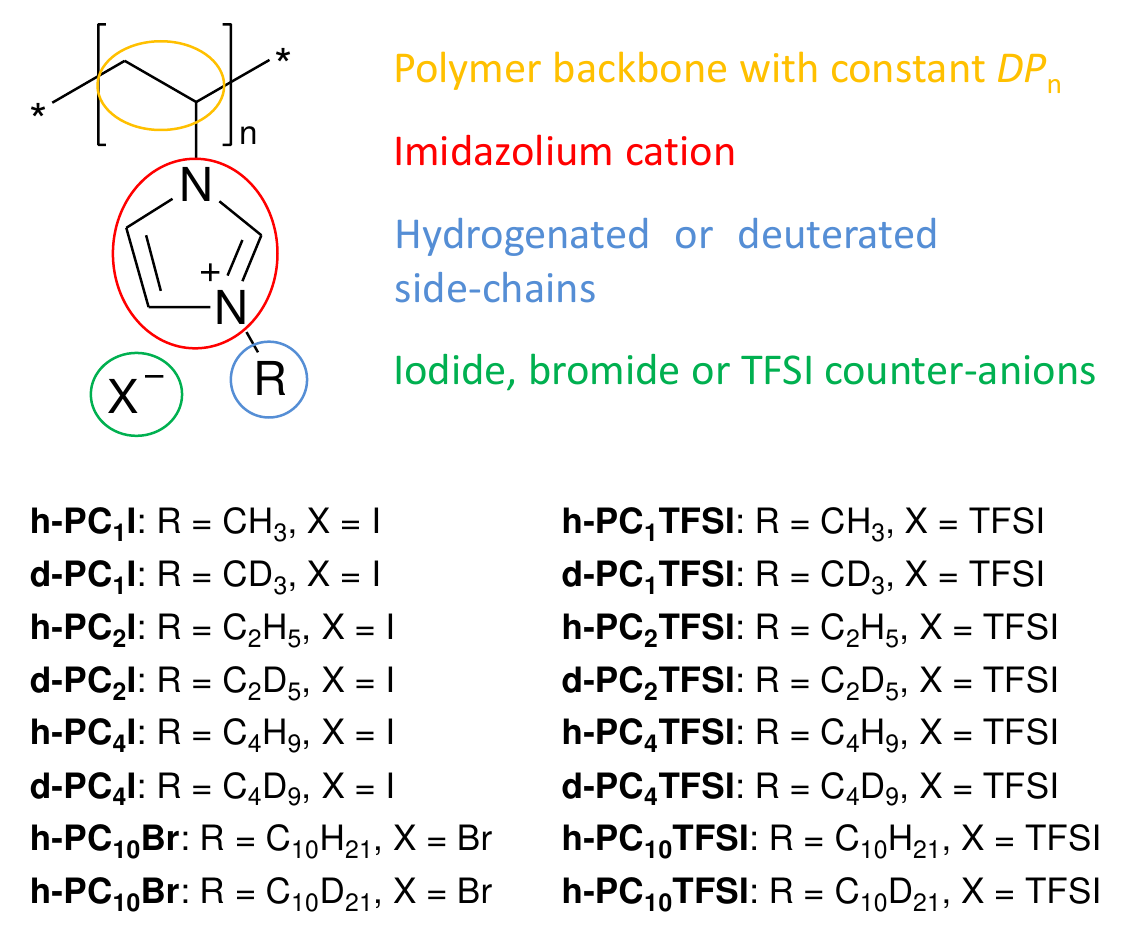}
	\caption{Chemical structure of hydrogenated and deuterated poly(1-vinyl-3-alkylimidazolium) isotopologues. \label{scheme}}
\end{scheme}

We investigate for the first time the influence of the alkyl side-chain length and the chemical nature of the counter-anion on the conformation of  poly(1-vinyl-3-alkylimidazolium) (\textbf{PC$_n$}) \cite{green_alkyl-substituted_2011} by means of an extensive use of Small Angle Neutron Scattering (SANS) which is the lone technique allowing to measure the form factor of a single polymer chain through selective deuteration \cite{cotton_conformation_1974}. Three cases are investigated here: dilute solutions of PILs in deuterated tetrahydrofuran (THF-d$_8$), dilute solutions in imidazolium ILs with identical alkyl substituents and finally, bulk PILs.

\section{Experiments}
\subsection{Materials}

A series of hydrogenated and deuterated poly(1-vinyl-3-alkylimidazolium) isotopologues having identical macromolecular parameters (\textit{i.e.} $N$ and \textit{Ð}) but including either halide or TFSI counter-anions and different N-3 side-chains of varying length (Scheme \ref{scheme}) were synthesized in three steps involving 1 - reversible addition-fragmentation chain transfer (RAFT) polymerization of vinyl imidazole (\textbf{VI}), 2 - N-alkylation of imidazole groups, and 3 - ion metathesis reaction (scheme in the Supplementary Information). Initially, \textbf{PVI} was obtained in 51 \% yield by RAFT polymerization of \textbf{VI} in methanol using a dithiocarbamate chain transfer agent (\textbf{CTA}) and \textbf{AIBN} as initiator. According to the initial $[\mathrm{\textbf{VI}}]/[\mathrm{\textbf{CTA}}]$ ratio of 139 and the monomer conversion obtained by \ce{^1H} NMR of the crude polymerization mixture (\textit{i.e.} 54 \%), the resulting polymerization degree ($N$) of \textbf{PVI} (and thus all their N-alkylated PIL derivatives) should be ca. 75.  A series of poly(1-vinyl-3-alkylimidazolium) halides was then obtained in 73 \% to 94 \% yields by $N$-alkylation of the N-3 position of imidazole groups using hydrogenated and deuterated aliphatic halide isotopologues with $n=$ 1, 2, 4 or 10 carbon atoms. Finally, all hydrogenated and deuterated poly(1-vinyl-3-alkylimidazolium) halides were involved in an ion metathesis reaction with LiTFSI to afford the corresponding poly(1-vinyl-3-alkylimidazolium) TFSI isotopologues in 81 \% to 98 \% yields. The detailed preparation and characterization of all PILs by \ce{^1H} NMR, \ce{^19F} NMR and SEC are described in the Supplementary Information.

\subsection{Small Angle Neutron Scattering}
For SANS experiments in bulk, thick films of PILs were prepared by solution casting of \textbf{PC$_n$TFSI} solutions in tetrahydrofuran. The drying of the solutions deposited on a 1 mm quartz plate was first performed at room temperature during 24 h in a dessicator with a small aperture to allow a slow evaporation rate preventing from crust / cracks formation. The films were then heated in an oven at 40 °C for 24 h in order to remove a maximum of remaining solvent without bubble formation. The third 24h annealing step is performed at 70 °C. Finally, the sample are heated at 100 °C under vacuum for 24 h. The thickness of the films was determined using a Palmer by averaging 10 measurements over the film area. The resulting films are 10 mm in diameter and around 0.8 mm thick. The same procedure was used to prepare \textbf{PC$_n$I} and \textbf{PC$_{10}$Br} films using N,N-dimethylformamide as the casting solvent.

To study PILs conformation in THF, in which \textbf{PC$_n$TFSI} exhibited very good solubility, hydrogenated \textbf{PC$_n$TFSI} were diluted  at 0.5 w\% in deuterated THF (THF-d8). Assuming that THF is a good solvent, the overlapping volume fraction can be estimated by \cite{rubinstein_polymer_2003}: $\phi^*=N^{-4/5} \approx$ 3\% which is well above our minimal concentration. Semi-dilute solutions of hydrogenated \textbf{PC$_4$TFSI} (1.5, 4.2 and 11 w\%) and \textbf{PC$_{10}$TFSI} (1.9, 3.8 and 9.8 w\%) were also prepared. The solutions were stored in 2 mm thick quartz cells.

For solutions of PILs in ILs, we used a concentration of 1 w\% of deuterated PILs in hydrogenated ILs (chemical structures are given in the Supplementary Information). Prior to mixing, ILs were dryed under vaccuum during 15 h. The quartz cell thickness was 1 mm in this case to reduce incoherent background and optimize transmission.
Small Angle Neutron Scattering (SANS) was performed on the small angle diffractometer PAXY at Laboratoire Léon Brillouin, CEA Saclay. Three different sample-to-detector distances of 1, 3 and 5 m were used together with a neutron wavelength of 4, 5 and 8.5 Å respectively. The available range of scattering vector Q=$\frac{4\pi}{\lambda}\sin(\theta)$ was thus 0.006 \AA$^{-1}$ < $Q$ < 0.6 \AA$^{-1} $.  Standard corrections were applied for sample volume, neutron beam transmission, empty cell signal, and detector efficiency to the raw signal to obtain scattering spectra in absolute units \cite{brulet_improvement_2007}.

\subsection{Wide Angle X-Ray Scattering}
Wide Angle X-Ray Scattering (WAXS) measurements were performed on the same films of hydrogenated PILs \textbf{h-PC$_n$} measured with SANS. The films were cut in 2 $\times$ 2 mm squares and placed on the sample holder. We used a Xeuss 2.0 diffractometer (Xenocs) equipped with a Pilatus 1M detector placed at 200 mm from the sample and a Genix 2D Cu HF source ($\lambda=1.542$ \AA). The same data correction scheme used for SANS was applied to WAXS spectra.  

\section{Results}
\subsection{PILs in THF Solutions}

\begin{table*}[ht]
	\begin{adjustbox}{width=2\columnwidth,center}
	\begin{tabular}{|c||c|c||c|c||c|c|c|}
		\hline
		& \multicolumn{2}{c||}{\textbf{Solution in THF}} & \multicolumn{2}{c||}{\textbf{Solution in IL}} & \multicolumn{3}{c|}{\textbf{Bulk}}\\
		\hline
		&&&&&&&\\[-1em]
		Sample & $R_g$ (\AA) & $r_{\mathrm{cross}}$ (\AA) & $R_g$ (\AA) & $r_{\mathrm{cross}}$ (\AA) & $R_g$ (\AA) & $r_{\mathrm{cross}}$ (\AA) & $d_{\mathrm{bb}}$ (\AA) \\
		&&&&&&&\\[-1em]
		\hline
		\textbf{PC$_1$TFSI} & - & - & - & - & - & - & - \\
		\textbf{PC$_2$TFSI} & 58.2 $\pm$ 2.1& 6.0 $\pm$ 0.6 & 41.8 $\pm$ 3.7 & 5.9 $\pm$ 295.3 & 41.9 $\pm$ 0.3 & 5.7 $\pm$ 0.1 & 11.9 $\pm$ 0.3 \\
		\textbf{PC$_4$TFSI} & 58.9 $\pm$ 1.7 & 7.7 $\pm$ 0.3 & 44.3 $\pm$ 1.3 & 5.6 $\pm$ 300.8 & 44.5 $\pm$ 0.4 & 7.1 $\pm$ 0.1 & 12.6 $\pm$ 0.1 \\
		\textbf{PC$_{10}$TFSI} & 64.4 $\pm$ 1.4& 10.7 $\pm$ 0.2 & 52.6 $\pm$ 0.7 & 16.5 $\pm$ 0.7 & 48.0 $\pm$ 0.2 & 12.6 $\pm$ 0.0& 20 $\pm$ 0.1 \\
		\hline
		
		\textbf{PC$_1$I} & - & - & - & - & 46.3 $\pm$ 0.7 & - & - \\

		\textbf{PC$_2$I} & - & - & - & - & 41.9 $\pm$ 0.3 & 5.7 $\pm$ 0.1& 11.9 $\pm$ 0.3 \\

		\textbf{PC$_4$I} & - & - & - & - & 39.7 $\pm$ 0.2 & 8.6 $\pm$ 0.0 & 13.6 $\pm$ 0.1 \\

		\textbf{PC$_{10}$I} & - & - & - & - & 38.1 $\pm$ 0.2 & 15.0 $\pm$ 0.0 & 25.6 $\pm$ 0.0 \\
		\hline
	\end{tabular}
	\caption{Fitting parameters of the SANS data. In solution in THF and IL, concentrations are respectively 0.5 w\% and 1 w\% in PIL.}
	\label{tablefit}
	\end{adjustbox}
\end{table*}

\begin{figure}
	\includegraphics[width=\columnwidth]{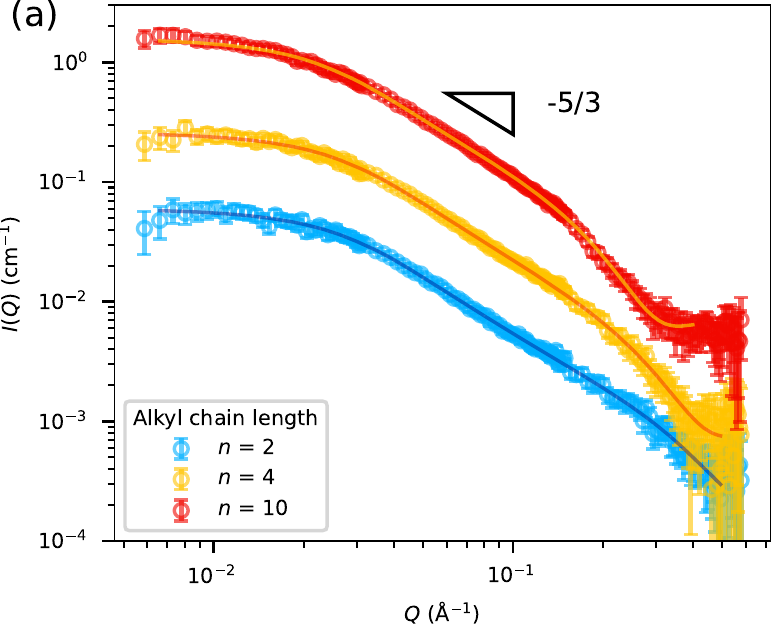}\par
	\includegraphics[width=\columnwidth]{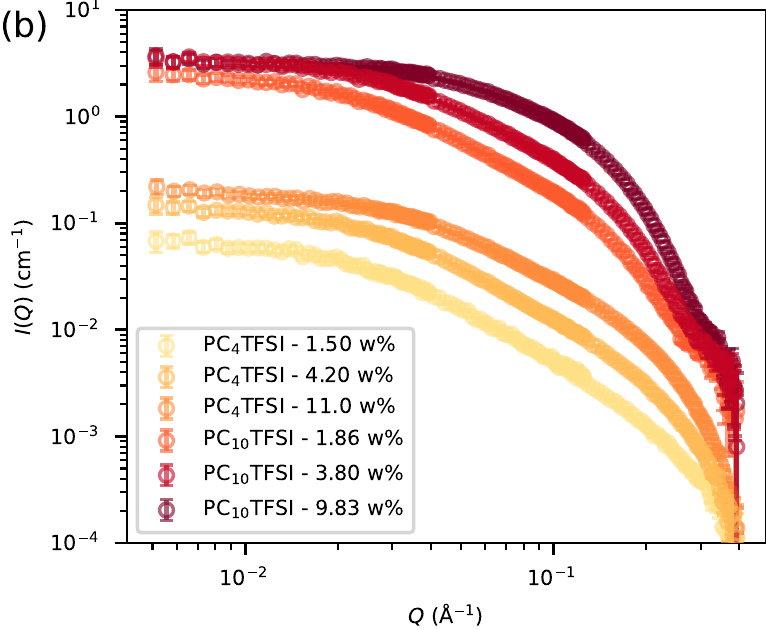}\par
	\caption{SANS spectra of \textbf{PC$_n$TFSI} solutions in THF-d$_8$ for different alkyl side-chain length $n$. (a) Dilute regime 0.5 w\%, 0.55 w\%, 0.5 w\% respectively for $n=2, 4, 10$. Solid lines correspond to the best fits according to equation \ref{eq_thf}. (b) Semi-dilute regime with varying concentrations for alkyl chain lengths $n=4$ and $10$.  Data are shifted vertically for clarity.}
	\label{sans_diluted}
\end{figure}

The SANS spectra of  diluted poly(1-vinyl-3-alkylimidazolium)s with TFSI counter-anions presented in Figure \ref{sans_diluted}-a show several interesting features. For all alkyl side-chain lengths studied, a clear transition from the Guinier to the intermediate regime allows the radius of gyration ($R_{\mathrm{g}}$) of the chains to be quantified. The intermediate regime scales as $Q^{-5/3}$, a fractal dimension in good agreement with chains in good solvent \cite{cotton_conformation_1974, de_gennes_p._g._scaling_1979} independantly of the alkyl side-chain length $n$. For higher values of the scattering vector $Q$, the SANS spectra exhibit a change in slope which onset depends on $n$. The larger $n$, the lower the $Q$ value at which this change in slope appears. Such feature is most likely due to the cross section of the PIL chains that understandably depends on $n$. In order to account for both $R_{\mathrm{g}}$ and the cross section of PILs ($r_{\mathrm{cross}}$), the data were fitted using an excluded volume polymer chain form factor $P_{\mathrm{excl}}(Q)$ and a cross section term $P_{\mathrm{cross}}(Q)$ from a rigid rod \cite{chen2006_cross_formfactor}. The contour length of the chains being significantly larger than the cross section, their respective scatterings can be separated by means of a decoupling approximation \cite{jerke1998_decouplingapprox}, giving:
\begin{equation}
\label{eq_thf}
\begin{aligned}
P(Q) &= P_{\mathrm{excl}}(Q)\times P_{\mathrm{cross}}(Q) \\
P_{\mathrm{excl}}(Q) &= 2\int_0^1(1-x)\exp{\left(-\frac{q^2 a^2}{6}N^{2\nu}x^{2\nu}\right)}\\
P_{\mathrm{cross}}(Q) &=\left(2 \frac{J_1(Qr_\mathrm{cross})}{Qr_\mathrm{cross}}\right)^2
\end{aligned}
\end{equation}
%
where $\nu=5/3$ is the excluded volume parameter. Regarding the cross section term, $J_1$ denotes the first order Bessel function of the first kind.
Such model proved to give a very accurate description of the scattering curve for all alkyl chain lengths studied here (Figure \ref{sans_diluted}). Relevant fitting parameters are given in Table \ref{tablefit}.

SANS spectra of semi-dilute solutions of \textbf{PC$_n$TFSI} in THF are also reported in Figure \ref{sans_diluted}-b. The absence of polyelectrolyte peak, assigned to intermolecular electrostatic interactions between charged chains \cite{combet2018polyelectrolytes} is noteworthy. A scaling of the intermediate regime as $Q^{-1}$ is also expected in this type of system when couter-ions are solubilized and the charged polyelectrolyte chain stretches into a rod-like structure under repulsive electrostatic interactions between monomer units \cite{degennes1976polyelectrolytes}. Together with the absence of polyelectrolyte peak, this represents clear evidence of at least partial condensation of the counter-anions on the PIL chains in solution in THF which may result from the low dielectric constant of THF and the strong hydrophobicity of TFSI counter-anions. Additionally, a similar change in slope at high $Q$ reveals the influence of the chain cross section.

\subsection{PILs in IL Solutions}
\begin{figure}
	\includegraphics[width=\linewidth]{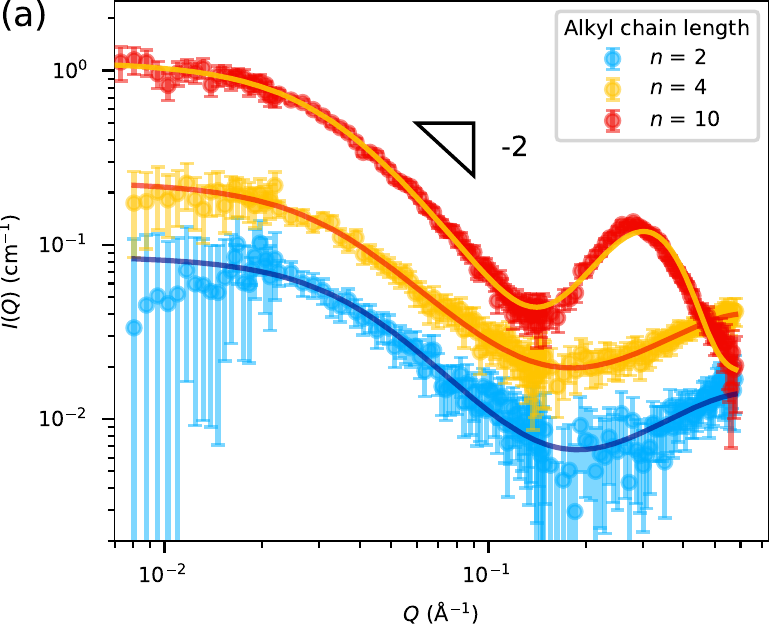}\par
	\includegraphics[width=\linewidth]{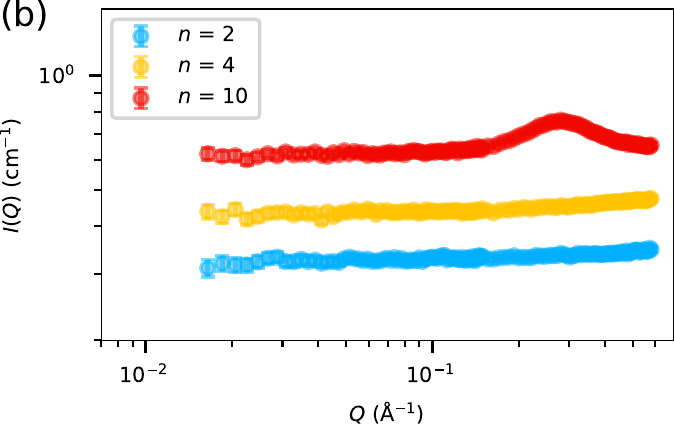}\par
	\caption{SANS spectra of (a) \textbf{PC$_n$TFSI} with different alkyl side-chain length $n$ in solution in the corresponding hydrogenated IL (\textit{i.e.} with the same $n$ and counter-anion) at a concentration of PIL of 1 w\% and (b) hydrogenated 1-alkyl-3-methylimidazolium ILs (with $n=2, 4, 10$ and \ce{TFSI^-} counter-anion). \label{sans_IL}}
\end{figure}

When deuterated PILs are diluted in a hydrogenated 1-alkyl-3-methylimodazolium IL with the same alkyl substituent and the same counter-anion, the SANS spectra (Figure \ref{sans_IL}-a) display an intermediate regime scaling as $Q^{-2}$. Hence the corresponding ILs behave as near $\Theta$ solvents. This can be understood by the similarity between monomer-monomer and monomer-solvent interactions.  The pronounced noise in the case of $n=2$ is due to the low contrast inherent from the decreasing proportion of deuterium with decreasing $n$. Additionally, we observe an upturn of the intensity at large $Q$ for $n=2$ and $n=4$ and a correlation peak for $n=10$. Such additional feature arises from the scattering signal of the IL itself \cite{hardacre_small_2010, annapureddy2010origin}. The well defined correlation peak observed for $n=10$ is a consequence of the amphiphilic nature of this IL leading  to the existence of structural heterogeneities \cite{russina_morphology_2009}. We chose not to subtract the volume weighted scattering from the solvent due to the little variation in water content of the latter that can cause the incoherent scattering background to vary between pure IL and the IL solvent in solution. Furthermore, even though we look at dilute solutions, addition of PILs can alter the structure of ILs \cite{widegren_effect_2005}. Here, PILs can be viewed as impurities within ILs modifying the structure of ILs but stating this point clearly deserves further investigations. By restraining the analysis to the lower $Q$ values of the spectra in which ILs alone do not coherently scatter (Figure \ref{sans_IL}-b), conformational properties of PIL chains can still be retrieved.
As a consequence, a good candidate to model the scattering spectra of dilute solutions of PILs in ILs can be infered from equation \ref{eq_thf} by replacing the excluded volume polymer chain form factor by a Debye model $P_\mathrm{Debye}(Q)$ for ideal chains \cite{debye_molecular-weight_1947, wiley_characterization_2008}. The fitting model can then be written as follows:
\begin{multline}
	\label{eq_melt}
	I(Q)=I_0 P_{\mathrm{Debye}}(Q)\times P_{\mathrm{cross}}(Q)+ \\	I_1\exp{\left(-\frac{(Q-Q_0)^2}{2 \sigma^2}\right)}
\end{multline}
where the gaussian function describes the contribution from the IL. As shown in Figure \ref{sans_diluted}-b and Table \ref{tablefit}, such model allows to get a good description of the experimental data but with a rather low precision on the determination of $r_{\mathrm{cross}}$. This can be explained by the lower signal-to-noise ratio at high scattering vectors compared to the previous case of deuterated THF and by the dominant contribution of the solvent at high $Q$. The determination of $R_{\mathrm{g}}$ remains nevertheless accurate.

\subsection{Bulk PILs}
\begin{figure}[h]
	\includegraphics[width=\linewidth]{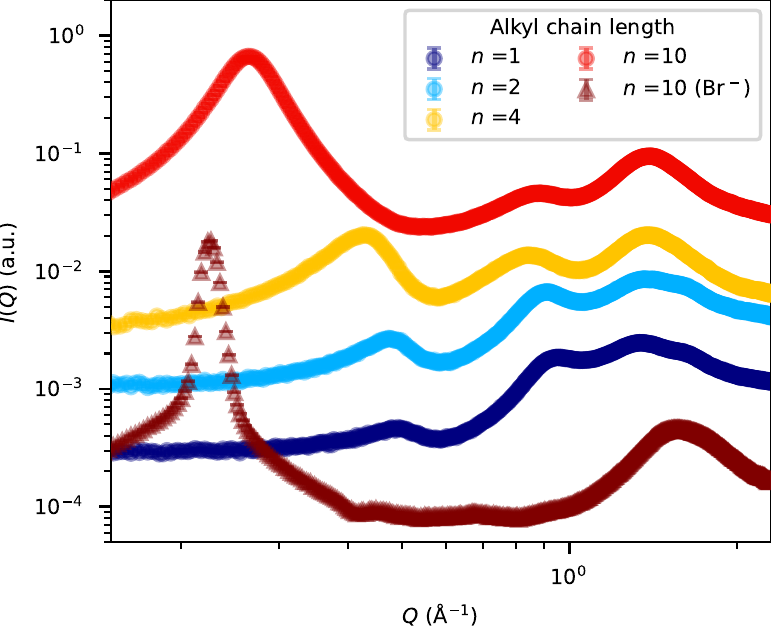}\par
	\caption{WAXS spectra of bulk \textbf{PC$_n$TFSI} ($\circ$) with different alkyl side-chain length $n$ and \textbf{PC$_{10}$Br} ($\bigtriangleup$). Data are shifted vertically for clarity. \label{waxs}}
\end{figure}

The WAXS patterns of bulk \textbf{PC$_n$TFSI} films, reported in Figure \ref{waxs} for several alkyl side-chain lengths $n$, feature three peaks in the $Q$ range 0.1 \AA$^{-1}$ < $Q$ < 2.5 \AA$^{-1}$. Previous works attributed these peaks to three specific correlation lengths of the system based on either simulated or experimental data and using selective deuteration to disentangle the contributions of the different characteristic lengths of the system \cite{liu_alkyl_2017, cruz_correlating_2012, araque2015modern, arbe2008anomalous}. The low-$Q$ peak (below 0.5 \AA$^{-1}$) is highly dependent on $n$ and shifts to low-$Q$ values while sharpening and gaining in intensity as $n$ increases. It is assigned to the backbone-to-backbone correlation length $d_{\textrm{bb}}$, in other words the distance between two neighboring macromolecular chains \cite{liu_direct_2016, iacob_polymerized_2017}. This distance increases as $n$ grows larger due to progressively larger alkyl domains. The increase in intensity stems from both a larger organized nanodomain and increasing contrast. The high-$Q$ peak is a common feature of X-ray scattering from molecular liquids and is attributed to close contact between alkyl side-chains inside the alkyl domain \cite{arbe2008anomalous, cruz_correlating_2012}. As for the intermediate peak (slightly below 1 \AA$^{-1}$), it is ascribed to close contact between same charges, \textit{i.e.} the correlation length of the counter-anion network. This last peak shows very little dependence on $n$. The ionic and pendant peaks will not be addressed in the following discussion. As for the WAXS of \textbf{PC$_n$I}, the high electron density of the iodide counter-anion hinders any kind of interpretation because of high X-Ray absorption (data not shown).

\begin{figure}
	\includegraphics[width=\linewidth]{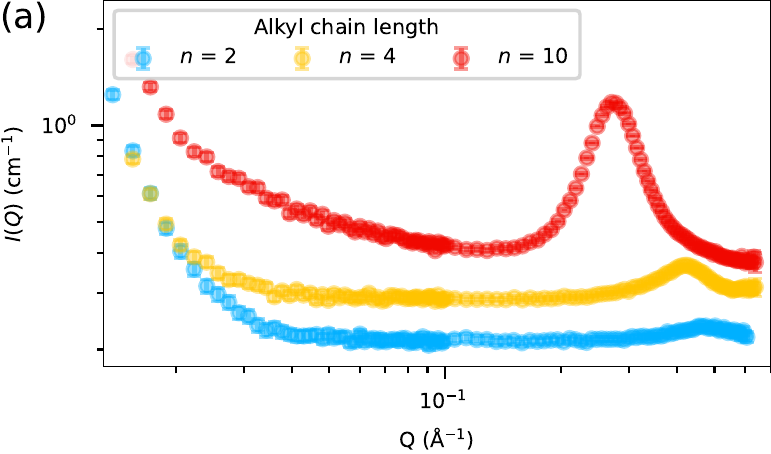}\par
	\includegraphics[width=\linewidth]{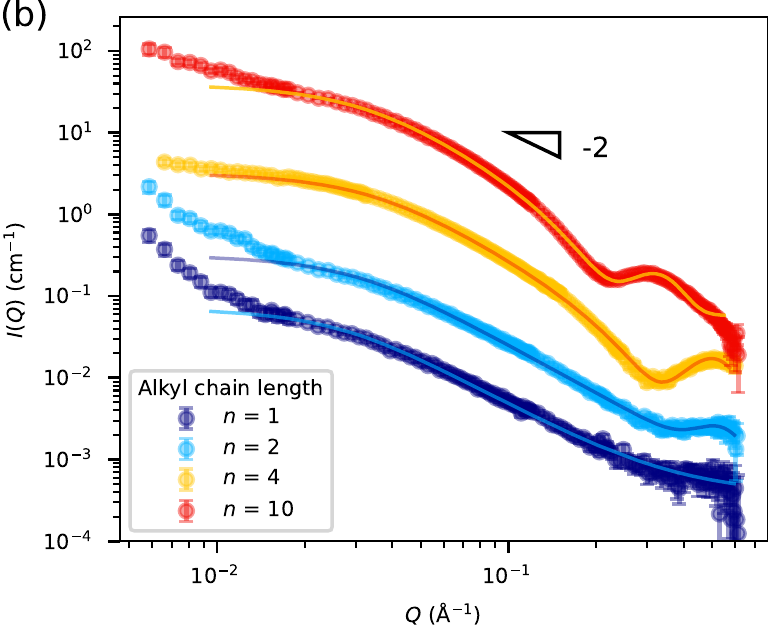}\par
	\includegraphics[width=\linewidth]{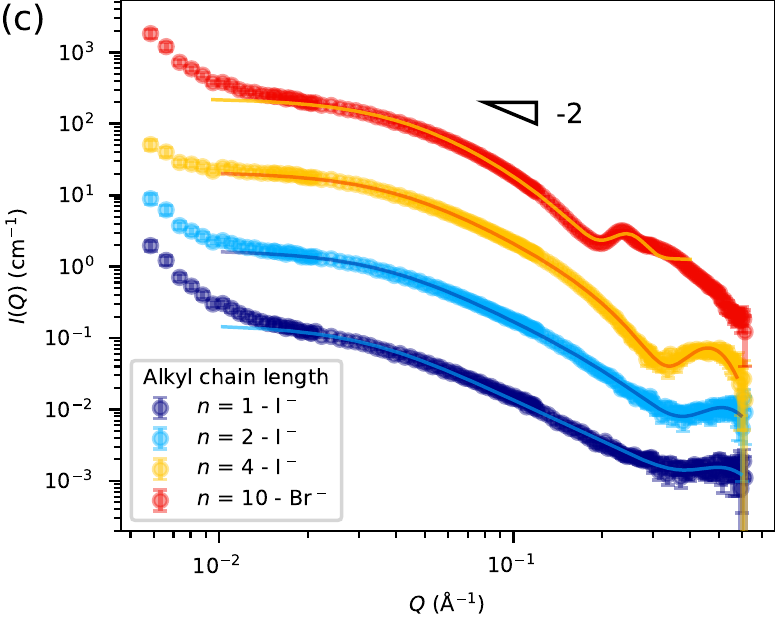}
	\caption{SANS spectra of bulk (a) purely deuterated \textbf{PC$_n$TFSI}, (b) \textbf{PC$_n$TFSI} H/D (1:1 w/w), and (c) bulk \textbf{PC$_n$I} and \textbf{PC$_{10}$Br} H/D (1:1 w/w) with varying alkyl side-chain length $n$. Solid lines correspond to the best fits according to equation \ref{eq_melt}. Data are shifted vertically for clarity. \label{sans_melt}}
\end{figure}

Many works focused on the very local structuration of PILs in bulk, whether using simulated neutron scattering \cite{liu_alkyl_2017,liu_direct_2016, heres_probing_2016}  or experimental X-Ray scattering\cite{doughty_structural_2019, cruz_correlating_2012, iacob_polymerized_2017}. Yet, to the best of our knowledge, the macromolecular conformation of PIL chains in bulk has not been investigated so far. SANS is the lone technique allowing to extract such information. When the sample is only composed of hydrogenated (h-PILs) or deuterated (d-PILs) PIL chains, SANS spectra can reveal their very local structure through a correlation peak (Figure \ref{sans_melt}-a), which is the same as the low-$Q$ peak of the WAXS patterns (Figure \ref{waxs}) corresponding to the backbone-to-backbone correlation length. An increase in intensity at low $Q$ values is also observed and is attributed to the presence of micro cracks in the samples inherent to the drying step in the sample preparation and which scatter at very low angle. Since h-PILs and d-PILs have identical $N$ values, it can be derived that the scattering intensity of a mixture of both is directly proportional to the form factor of PIL chains $P(Q)$ \cite{guinier_small-angle_1955}. Figures \ref{sans_melt}-b and \ref{sans_melt}-b respectively present the SANS spectra for H/D \textbf{PC$_n$TFSI} and \textbf{PC$_n$I}/\textbf{Br} in bulk for $n=$1, 2, 4 and 10. Similarly to the fully deuterated samples, we observe a slight increase at very low $Q$ which also stems from micro cracks in the sample. This undesired feature is then followed by a Guinier regime and an intermediate regime scaling as $Q^{-2}$, characteristic of an ideal polymer chain. For higher $Q$ values, the results show two additional features: a change in slope and a correlation peak. Similarly to the SANS spectra in THF (Figure \ref{sans_diluted}-a), for $n\geq4$, the change in slope ascribed to the contribution of the chain's cross section is observed. All in all, the same model as equation \ref{eq_melt} gives a very satisfying description of the scattering spectra of PILs in bulk with a peak function describing the backbone-to-backbone distance $d_\mathrm{bb}=\frac{2\pi}{Q_0}$ (Figure \ref{sans_melt}-b and \ref{sans_melt}-c). The cross section and $d_{\mathrm{bb}}$ of \textbf{PC$_1$TFSI} was too small to be estimated correctly with this fitting model. The nature of the counter-anions do not change the intrinsic conformation of the PIL chains, thus the scattering curves are very similar at first glance. However, fitting of the data does unveil an effect on parameters such as the radius of gyration $R_g$ as will be discussed hereafter. It must be noted at this point that the correlation peak for \textbf{PC$_{10}$Br} is of peculiar shape compared to other PILs. This is due to partial crystallization of the material, supported by the very sharp low-$Q$ peak on the WAXS signal (Figure \ref{waxs}).

\section{Discussion}
\begin{figure}
	\includegraphics[width=\linewidth]{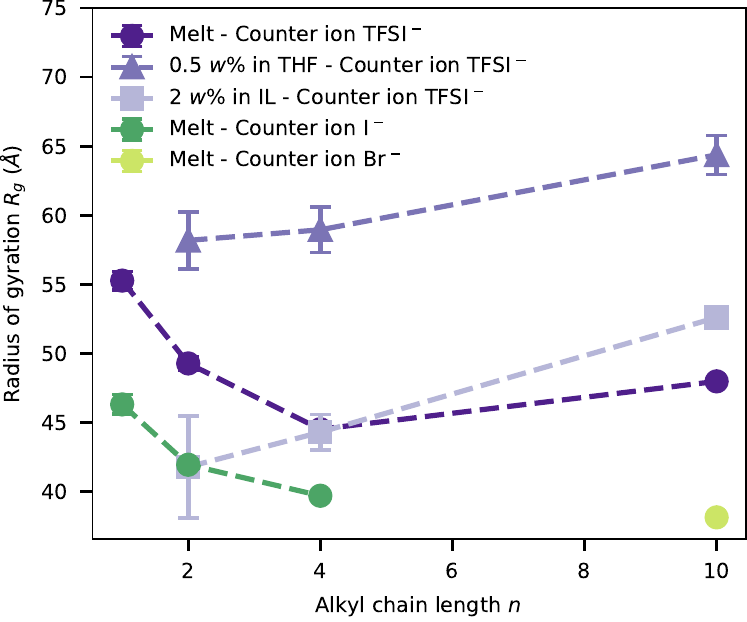}
	\caption{Radii of gyration of \textbf{PC$_n$TFSI} as a function of $n$, diluted in THF-d$_8$ (\textcolor{purpleTHF}{$\blacktriangle$}), in 1-alkyl-3-methylimidazolium ionic liquids with identical alkyl groups (\textcolor{purpleIL}{$\blacksquare$}) and of bulk \textbf{PC$_n$} with counter-anions TFSI (\textcolor{purpleMelt}{$\bullet$}), I$^{-}$ (\textcolor{green}{$\bullet$}) and Br$^{-}$ (\textcolor{lightgreen}{$\bullet$}). The dashed lines are eyes guideline. \label{fig:rg}}
\end{figure}

From the modeling of the SANS spectra, we follow the influence of the alkyl chain length $n$ on the radius of gyration, $R_g$, the cross section $r_\mathrm{cross}$ and the backbone-to-backbone distance $d_\mathrm{bb}$ in the three cases of interest. The evolution of $R_g$ is reported in Figure \ref{fig:rg} and shows an increasing behavior with $n$ for PIL chains diluted in both THF and ILs. PIL chains are larger in THF because of its good solvent property. As the size of the monomer increases, the Khun length increases because of local steric effect originating from the side-chains which leads to an overall increase of the chain's coil size. The most astonishing effect lies in the evolution of $R_{\mathrm{g}}$ in bulk, for which the coil size first decreases until $n=4$ and then slightly increases for $n=10$. Such non monotonic evolution goes against a simple steric effect of the side-chains since an increasing crowding of the monomer induces a progressive collapse of the coil for short side-chains. The same decrease of $R_g$ is observed for iodide counter-anions with a vertical shift to lower values of radii due to the smaller size of the ion compared to TFSI. The following increase cannot be readily confirmed by our experiments but considering the radius of gyration for the bromide counter-anion, again smaller than the iodide, seems to remain consistent with the non monotonic behavior described above. As macromolecular parameters (\textit{i.e.} $N$ and \textit{Ð}) of all the samples are strictly identical, this behavior can only be attributed to the length of the side-chain and may reflect a change in the flexibility of PILs' backbone. Atomistic Molecular Dynamics Simulations performed by Liu \textit{et al.} \cite{liu_direct_2016,liu_alkyl_2017} on poly(1-vinyl-3-alkylimidazolium) with TFSI counter-anions have shown that longer alkyl chains are more flexible and can form a bicontinuous sponge like nanostructure similar to ILs for $n \geqslant5$. Still, the lone flexibility of the side-chain does not necessarily explain the overall decrease of the backbone flexibility. Analysis of WAXS signals from the same samples (Figure \ref{waxs}) show that the backbone-to-backbone distance increases by $\approx 1.26$~\AA ~ per carbon which is very consistent with the work of Iacob \textit{et al.}\cite{iacob_polymerized_2017} on poly(1-vinyl-3-alkylimidazolium)s with varying counter-anions (including TFSI) and alkyl chain lengths $n$ comprised between 2 and 6, obtained by free radical polymerization (\textit{i.e.} comparable $N$ but higher \textit{Ð}). Such increase is slower than expected by assuming that pendant groups do not interdigitate\cite{delhorbe_unveiling_2017, cruz_correlating_2012, liu_alkyl_2017, liu_direct_2016} thus giving a first suggestion of the interpenetration of long alkyl chains.

\begin{figure}
	\includegraphics[width=\linewidth]{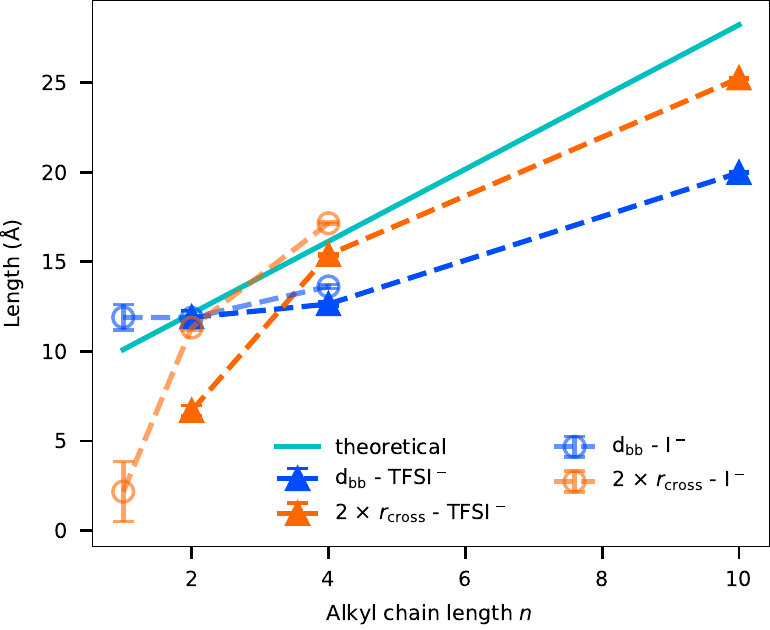}
	\caption{Evolution of $2\times r_{\mathrm{cross}}$ (\textcolor{BurntOrange}{$\blacktriangle$}) and $d_\mathrm{bb}$ (\textcolor{NavyBlue}{$\blacktriangle$}) of bulk \textbf{PC$_n$TFSI} and \textbf{PC$_n$I} (\textcolor{BurntOrange}{$\circ$}, \textcolor{NavyBlue}{$\circ$}) as a function of the alkyl chain length $n$. The solid line represents the theoretical chain diameter. The dashed lines are eyes guideline.}
	\label{fig:cross}
\end{figure}

The above SANS experiments can also probe both the backbone-to-backbone distance and the chain's cross section, only estimated in previous works either by MD simulations or density functional theory (DFT) calculations in the scope of the same kind of comparison \cite{delhorbe_unveiling_2017, heres_probing_2016, doughty_structural_2019}. Figure \ref{fig:cross} represents the influence of the alkyl chain length on both the diameter ($2r_\mathrm{cross}$) of bulk PIL chains and the backbone-to-backbone distance for both TFSI and iodide counter-anion. The increase in $2r_\mathrm{cross}$ appears to be more sensitive to $n$ than $d_\mathrm{bb}$. Moreover, for $n\geq 4$ the backbone-to-backbone distance becomes smaller than the chain diameter. At this point, side-chains could either bend or indeed interdigitate as suggested by the WAXS data. It was shown that even though alkyl chains gain in flexibility as they become longer \cite{liu_alkyl_2017}, they remain stiffer than an ether based side-chain \cite{doughty_structural_2019} for example, thus they are less likely to bend.

This is a direct proof that pendant chains tend to interpenetrate as their length increases and allows us to suggest a description of the molecular mechanisms responsible of the non monotonic evolution of $R_\mathrm{g}$ in Figure \ref{fig:rg}. For small $n$, the side-chains are relatively stiff and do not interdigitate which results in an overall low chain flexibility. As $n$ increases, a competition between the steric crowding between neighboring alkyl chains along the backbone tends to increase the Khun length as observed in dilute solutions (Figure \ref{sans_diluted}). Also, an increase in side-chain flexibility and interdigitation makes the chain more flexible, this being dominant for large values of $n$ according to our experimental data. In such a scenario, the contribution of the electrostatic interactions between monomers is not taken into account. They may also play an important role at small scales and are highly dependent on the degree of ion pairing as proposed by  Gebbie \textit{et al.} \cite{agebbie_long_2017} for ILs using Surface Force Apparatus experiments. Increasing the length of the alkyl side-chains can play a significant role in the screening of electrostatic interactions between cations also resulting in an increase of chain flexibility. However, a direct comparison between SFA experiment on ILs and our results may be hazardous because of the introduction of intra- and inter-chain Coulombic interactions which may differ because of the reduction of degree of freedom of the IL units after polymerization. Here, the probed radii of gyration only depend on the intra-chain interaction as the scattering signal from a mixture of hydrogenated and deuterated chains cancels the inter-chain terms of the partial structure factors\cite{cotton_conformation_1974}. Elucidating such difference may be of great interest but remains challenging since it requires to uncouple steric and electrostatic interactions.

\section{Conclusion}
In summary, an extensive use of small angle neutron scattering provides a multiscale description of the conformation of poly(1-vinyl-3-alkylimidazolium)s in dilute solutions in good solvent, $\Theta$ solvent and in bulk as a function of the side-chain length and counter-anion chemical structure. The modelling of the scattering curves supported by physical arguments provided a quantitative measurement of both the radii of gyration $R_g$, the chain cross sections $r_{\textrm{cross}}$ and the backbone-to-backbone distances $d_{\textrm{bb}}$. The overall results have shown a monotonic increase in $R_g$ as a function of $n$ for PILs in dilute solution, in agreement with increasing steric interactions between monomer units. In bulk, a non monotonic increase of $R_g$ of PIL chains appears and is interpreted as a competition between steric interactions, side-chain flexibility and potential electrostatic screening. Overall, the main chain  gains in flexibility, allowing the collapse of the coil before crowding effects take over for longer side-chains. Interpenetration of the latter was first suggested by the slow increase of backbone-to-backbone correlation length with $n$ and confirmed by a crossover between the cross section and the inter-chain distance, both of them estimated experimentally. Subsequently, the local interaction inherited from the IL monomer translates into a change of PILs conformation at the macromolecular scale which can be used to probe them. Relating these structural features to the understanding of the dynamic properties of PILs remains up to now an open question.

\begin{suppinfo}
\begin{itemize}
	\item Synthetic details, \ce{^1H} and \ce{^19F} NMR spectra of all PILs.
\end{itemize}
	
\end{suppinfo}

\begin{acknowledgement}
	This work was supported by ANR-POILLU program (ANR-19-CE06-0007). The authors thank Annie Brulet for fruitful discussions on Small Angle Neutron Scattering on polymer melts.
\end{acknowledgement}

%


\bibliographystyle{achemso}
\bibliography{biblio}


\end{document}


\newpage

\section{PILs Synthesis}
\begin{scheme}
	\includegraphics[width=\linewidth]{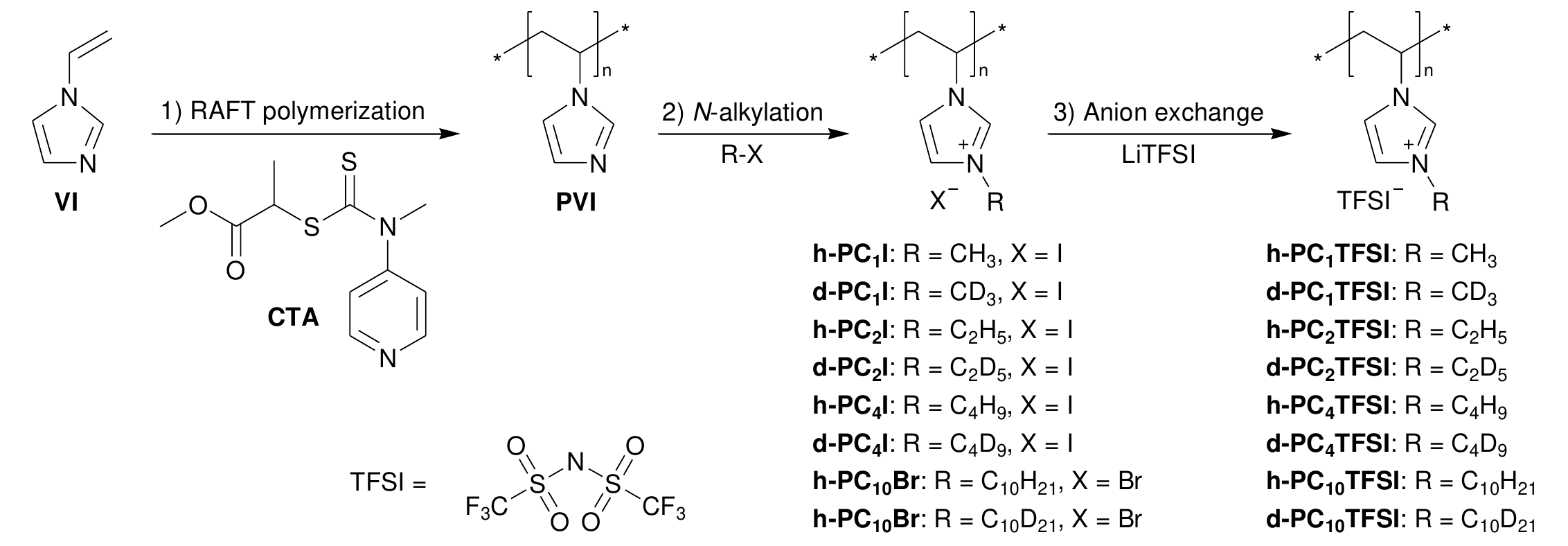}
	\caption{Synthesis of hydrogenated and deuterated poly(1-vinyl-3-alkylimidazolium) isotopologues. \label{fig:schemeS1}}
\end{scheme}

\subsection{Materials}
1-Vinylimidazole (\textbf{VI}, 99 \%), methyl 2-[methyl(4-pyridinyl)carbamothioylthio]propionate (\textbf{CTA}, 97 \%), 2,2'-azobis(2-methylpropionitrile) (AIBN, 98 \%), iodomethane (99 \%), iodoethane (99 \%) 1-iodobutane (99 \%), 1-bromodecane (98 \%), iodomethane-\ce{d_3} (99.5 \%), iodoethane-\ce{d_5} (99.5\%), lithium bis(trifluoromethylsulfonyl)imide (LiTFSI, 99.95 \%) were purchased from Merck and used as received. 1-Iodobutane-\ce{d_9} (98 \%) and 1-bromodecane-\ce{d_21} (98 \%) were purchased from CDN Isotopes and used as received. Deuterated tetrahydrofuran (THF-d$_8$ 99.5 \%) was purchased from Eurisitop. 1-Decyl-3-methylimidazolium bis(trifluoromethylsulfonyl)imide (99 \%) was purchased from Iolitec. 1-butyl-3-methylimidazolium (98 \%) and 1-ethyl-3-methylimidazolium (99 \%)  bis(trifluoro\-methylsulfonyl)imide were purchased from Sigma Aldrich. The ionic liquids were dried at 60 °C under vacuum for 24h before use.

\textbf{Nomenclature.} Poly(ionic liquid)s are designated as \textbf{Y-PC$_n$X} with X the nature of the counter anion (X = I, Br, or TFSI for iodide, bromide or bis(trifluoromethylsulfonyl)imide, respectively), n the number of carbon atoms of the N-3 alkyl side–chain ($n$ = 1, 2, 4 or 10 carbon atoms), and Y the isotopic nature of the N-3 alkyl side–chain (Y = h or d for perhydrogenated or perdeuterated isotopologues, respectively).

\begin{scheme}
	\includegraphics[width=0.6\linewidth]{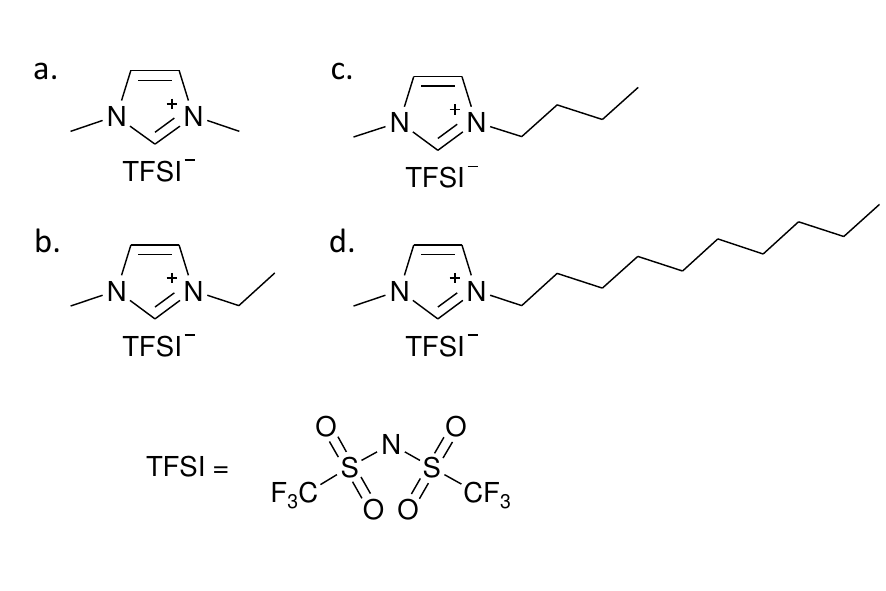}
	\caption{Chemical structure of  1-alkyl-3-methylimidazolium ILs (with alkyl chain lengths $n$ = 1 (a), 2 (b), 4 (c) and 10 (d)). \label{fig:schemeS2}}
\end{scheme}

\subsection{Characterization Methods}
\ce{^{1}H} (400 MHz) and \ce{^{19}F} (376.5 MHz) NMR data were recorded on a Bruker Avance 400 spectrometer in DMSO-\ce{d_6}. Chemical displacements ($\delta$) were listed with respect to the signal of residual internal \ce{CD_3SOCD2H} ($\delta= 2.50$) for \ce{^1H} spectra and to the signal of internal \ce{CFCl_3} ($\delta=0.00$) for \ce{^19F} spectra. Size exclusion chromatography (SEC) was carried out at 50 °C on a chromatograph connected to a Viscotek pump (1 mL min$^{-1}$) and Rheodyne 7725i manual injector (100 $\mu$L loop) using a combination of detectors (Viscotek VE3580 refractometer RI at 50 °C and Viscotek T60A viscometer at room temperature), two Viscotek I-MBHMW-3078 columns and one Viscotek I-MBLMW-3078 column, 300 × 7.5 mm (polystyrene/divinylbenzene) and pre-column Viscotek I-GUARD-0478 and a 0.01 M solution of LiTFSI in DMF as the eluent. 3 mg/mL solutions of \textbf{h-PC$_1$TFSI}, \textbf{h-PC$_2$TFSI}, \textbf{h-PC$_4$TFSI} and \textbf{h-PC$_{10}$TFSI} in 0.01 M LiTFSI in DMF were filtered through 0.20 $\mu$m pore size PTFE filter prior to the measurements. Number average (\textit{M}$_{\mathrm{n}}$) and weight average (\textit{M}$_{\mathrm{w}}$) molar masses and dispersities (\textit{Đ}) were derived from a calibration curve based on polystyrene standards. Omnisec software was used for the treatment of the results. 

\subsection{RAFT Polymerization of Vinyl Imidazole}
\begin{figure}
	\includegraphics[width=0.8\textwidth]{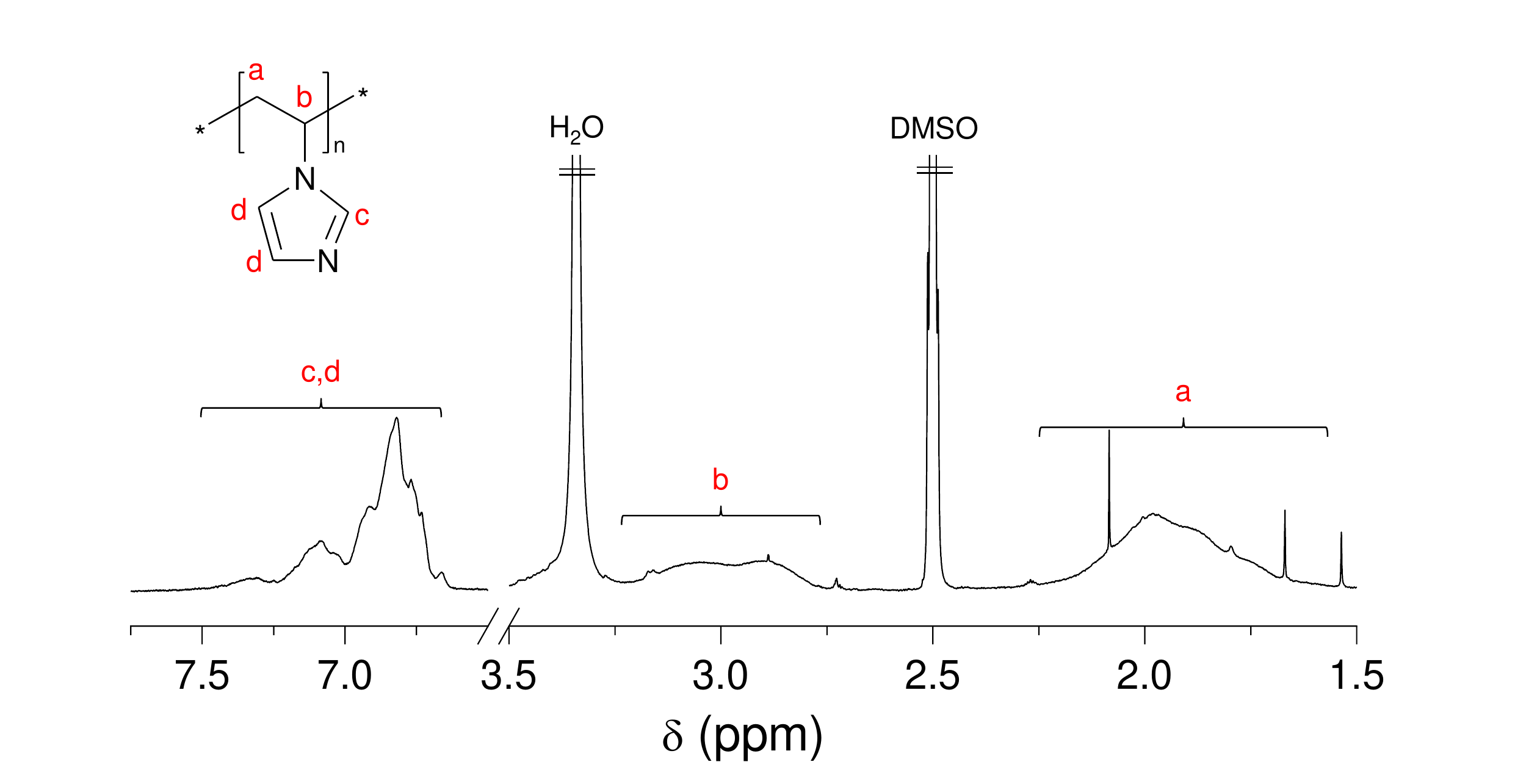}
	\caption{\ce{^{1}H} NMR spectrum (DMSO-\ce{d_6}, 400 MHz) of \textbf{PVI}.}
	\label{fig:nmr_PVI}
\end{figure}

Vinyl imidazole (7.70 g, 81.8 mmol) was added to a solution of \textbf{CTA} (0.16 g, 0.59 mmol) and AIBN (0.05 g, 0.30 mmol) in methanol (\ce{CH_3OH}, 30 mL). The solution was sealed under reduced pressure after three freeze-pump-thaw cycles and further stirred for 16 h at 70 °C. The crude solution was reduced under vacuum, precipitated in cold diethyl ether and dialyzed for 3 days in \ce{CH_3OH} to remove monomer traces and freeze-dried to afford \textbf{PVI} as a white solid (3.93 g, 51.0 \%). \ce{^1H} NMR (400 MHz, DMSO-\ce{d_6}): $\delta$ 7.57–6.59 (3H, br, \ce{H_c}, \ce{H_d}), 3.23–2.74 (1H, br, \ce{H_b}), 2.34–1.56 (2H, br, \ce{H_a}). \ce{^1H} NMR of \textbf{PVI} (Figure \ref{fig:nmr_PVI}) shows the quantitative removal of residual VI after purification (i.e. doublet of doublet at 7.17 ppm and doublets at 5.48 and 4.87 ppm) and the fair agreement between integrals of the imidazole protons (i.e. at 7.57–6.59 ppm) and the protons of the main–chain (i.e. at 3.23–2.74 and 2.34–1.56 ppm).

\subsection{General Procedure for $N$-alkylation Reaction} 
\textbf{Synthesis of h-PC$_1$I.} A solution of \textbf{PVI} (1.50 g, 15.9 mmol of imidazole groups) and iodomethane (22.6 g, 159 mmol) in DMF (60 mL) was heated at 80 °C for 24h. The resulting mixture was evaporated under reduced pressure, dissolved in acetonitrile and precipitated twice in cold acetone to afford after drying under reduced pressure \textbf{h-PC$_1$I} as a yellow solid (3.42 g, 91.1 \%). \ce{^1H} NMR (DMSO-\ce{d_6}, 400 MHz): $\delta$ 9.71–8.78 (1H, br, \ce{H_c}), 8.20–7.24 (2H, br, \ce{H_d}), 4.91–4.15 (1H, br, \ce{H_b}), 4.12–3.58 (3H, br, \ce{H_e}), 2.94–2.01 (2H, br, \ce{H_a}).

\textbf{Synthesis of d-PC$_1$I}. The general procedure for $N$-alkylation was applied to a mixture of \textbf{PVI} (795 mg, 8.44 mmol of imidazole groups) and iodomethane-\ce{d_3} (2.45 g, 16.9 mmol) to yield \textbf{d-PC$_1$I} as a white solid (1.85 g, 91.7 \%). \ce{^1H} NMR (DMSO-\ce{d_6}, 400 MHz): $\delta$ 9.64–8.76 (1H, br, \ce{H_c}), 8.25–7.25 (2H, br, \ce{H_c}), 4.95–4.03 (1H, br, \ce{H_b}), 3.04–1.98 (2H, br, \ce{H_a}).

\begin{figure}[H]
	\includegraphics[width=0.8\textwidth]{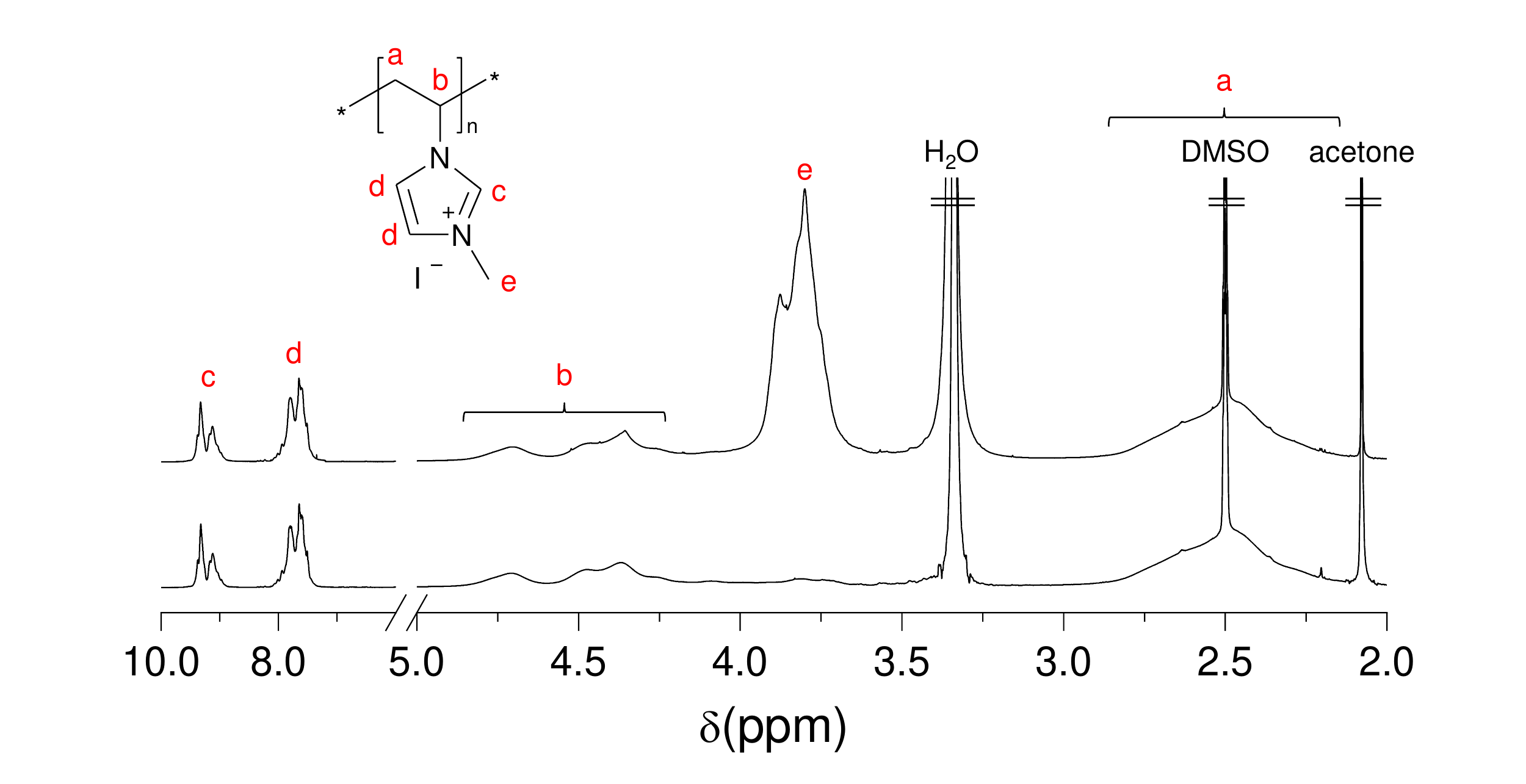}
	\caption{\ce{^{1}H} NMR spectrum (DMSO-\ce{d_6}, 400 MHz) of \textbf{h-PC$_1$I} (top) and \textbf{d-PC$_1$I} (bottom).}
	\label{fig:nmr_IH1_ID1}
\end{figure}
\newpage
\textbf{Synthesis of h-PC$_2$I.} The general procedure for $N$-alkylation was applied to a mixture of \textbf{PVI} (760 mg, 8.08 mmol of imidazole groups) and iodoethane (2.51 g, 16.1 mmol) to yield \textbf{h-PC$_2$I} as a light brown solid (1.78 g, 88.1 \%). \ce{^1H} NMR (DMSO-\ce{d_6}, 400 MHz): $\delta$ 9.68–8.94 (1H, br, \ce{H_c}), 8.22–7.11 (2H, br, \ce{H_d}), 4.89–3.70 (3H, br, \ce{H_b}, \ce{H_e}), 3.03–2.01 (2H, br, \ce{H_a}), 1.67–1.25 (3H, br, \ce{H_f}).

\textbf{Synthesis of d-PC$_2$I.} The general procedure for $N$-alkylation was applied to a mixture of \textbf{PVI} (744 mg, 7.91 mmol of imidazole groups) and iodoethane-\ce{d_5} (1.91 g, 11.9 mmol) to yield \textbf{d-PC$_2$I} as a brown solid (1.88 g, 93.2 \%). \ce{^1H} NMR (DMSO-\ce{d_6}, 400 MHz): $\delta$ 9.64–8.92 (1H, br, \ce{H_c}), 8.18–7.20 (2H, br, \ce{H_d}), 4.93–3.68 (1H, br, \ce{H_b}), 2.98–1.96 (2H, br, \ce{H_a}).

\begin{figure}[H]
	\includegraphics[width=0.8\textwidth]{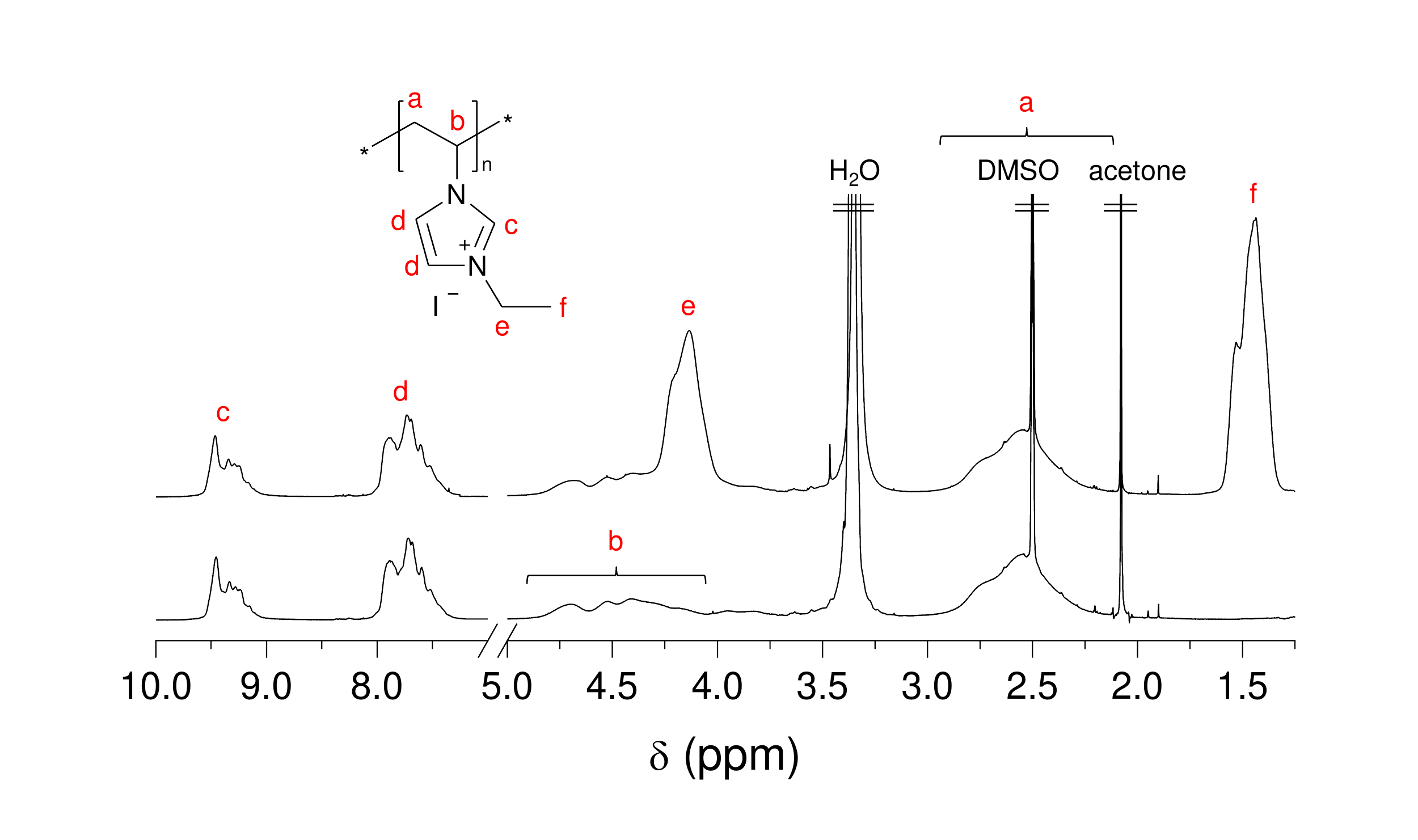}
	\caption{\ce{^{1}H} NMR spectrum (DMSO-\ce{d_6}, 400 MHz) of \textbf{h-PC$_2$I} (top) and \textbf{d-PC$_2$I} (bottom).}
	\label{fig:nmr_IH2_ID2}
\end{figure}
\newpage
\textbf{Synthesis of h-PC$_4$I.} The general procedure for $N$–alkylation was applied to a mixture of \textbf{PVI} (940 mg, 10.0 mmol of imidazole groups) and 1–iodobutane (2.76 g, 15.0 mmol) to yield \textbf{h-PC$_4$I} as a brown solid (2.56 g, 92.0 \%). \ce{^1H} NMR (DMSO-\ce{d_6}, 400 MHz): $\delta$ 9.80–8.94 (1H, br, \ce{H_c}), 8.16–7.19 (2H, br, \ce{H_d}), 4.95–3.73 (3H, br, \ce{H_b}, \ce{H_e}), 3.00–2.14 (2H, br, \ce{H_a}), 2.03–1.59 (2H, br, \ce{H_f}), 1.53–1.16 (2H, br, \ce{H_g}), 1.12–0.72 (3H, br, \ce{H_h}).

\textbf{Synthesis of d-PC$_4$I.} The general procedure for $N$-alkylation was applied to a mixture of \textbf{PVI} (910 mg, 9.67 mmol of imidazole groups) and 1-iodobutane-\ce{d_9} (2.80 g, 14.5 mmol) to yield \textbf{d-PC$_4$I} as a brown solid (2.02 g, 72.7 \%). \ce{^1H} NMR (DMSO-\ce{d_6}, 400 MHz): $\delta$ 9.78–8.93 (1H, br, \ce{H_c}), 8.17–7.16 (2H, br, \ce{H_d}), 5.89–3.72 (1H, br, \ce{H_b}), 3.05–2.12 (2H, br, \ce{H_a}).

\begin{figure}[H]
	\includegraphics[width=0.8\textwidth]{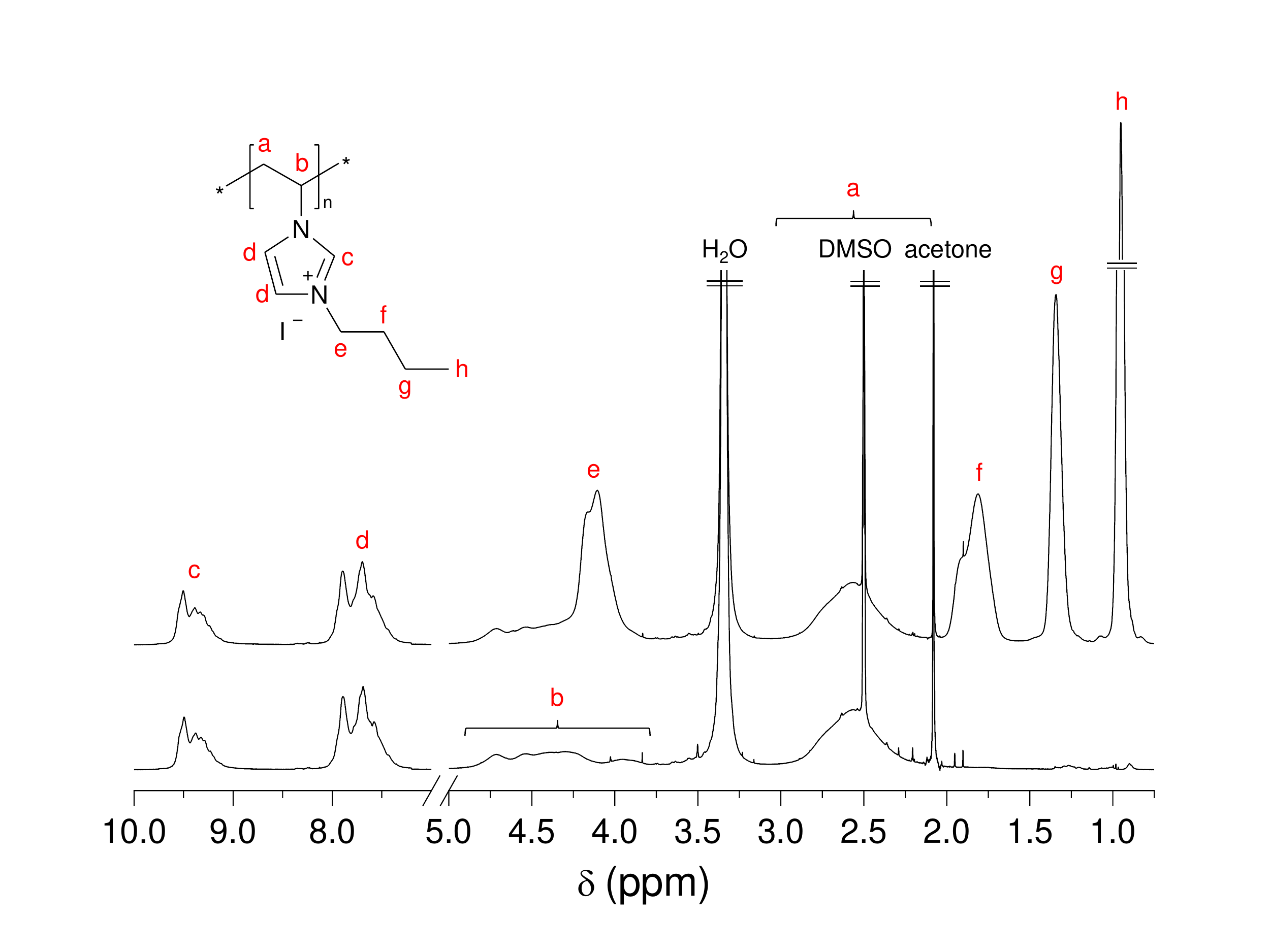}
	\caption{\ce{^{1}H} NMR spectrum (DMSO-\ce{d_6}, 400 MHz) of \textbf{h-PC$_4$I} (top) and \textbf{d-PC$_4$I} (bottom).}
	\label{fig:nmr_IH4_ID4}
\end{figure}
\newpage
\textbf{Synthesis of h-PC$_{10}$Br.} The general procedure for $N$-alkylation was applied to a mixture of \textbf{PVI} (602 mg, 6.40 mmol of imidazole groups) and 1-bromodecane (2.12 g, 9.58 mmol) to yield \textbf{h-PC$_{10}$Br} as a light yellow solid (1.89 g, 93.7 \%). \ce{^1H} NMR (DMSO-\ce{d_6}, 400 MHz): $\delta$ 10.17–9.16 (1H, br, \ce{H_c}), 8.54–7.25 (2H, br, \ce{H_d}), 5.06–3.70 (3H, br, \ce{H_b}, \ce{H_e}), 3.12–2.13 (2H, br, \ce{H_a}), 2.13–1.58 (2H, br, \ce{H_f}), 1.58–1.04 (14H, br, \ce{H_g}), 1.04–0.68 (3H, br, \ce{H_h}).

\textbf{Synthesis of d-PC$_{10}$Br.} The general procedure for $N$-alkylation was applied to a mixture of \textbf{PVI} (527 mg, 5.60 mmol of imidazole groups) and 1-bromodecane-\ce{d_21} (2.05 g, 8.46 mmol) to yield \textbf{d-PC$_{10}$Br} as a light yellow solid (1.76 g, 93.4 \%). \ce{^1H} NMR (DMSO-\ce{d_6}, 400 MHz): $\delta$ 10.20–9.03 (1H, br, \ce{H_c}), 8.46–7.28 (2H, br, \ce{H_d}), 5.08–3.84 (1H, br, \ce{H_b}), 3.11–2.13 (2H, br, \ce{H_a}).

\begin{figure}[H]
	\includegraphics[width=0.8\textwidth]{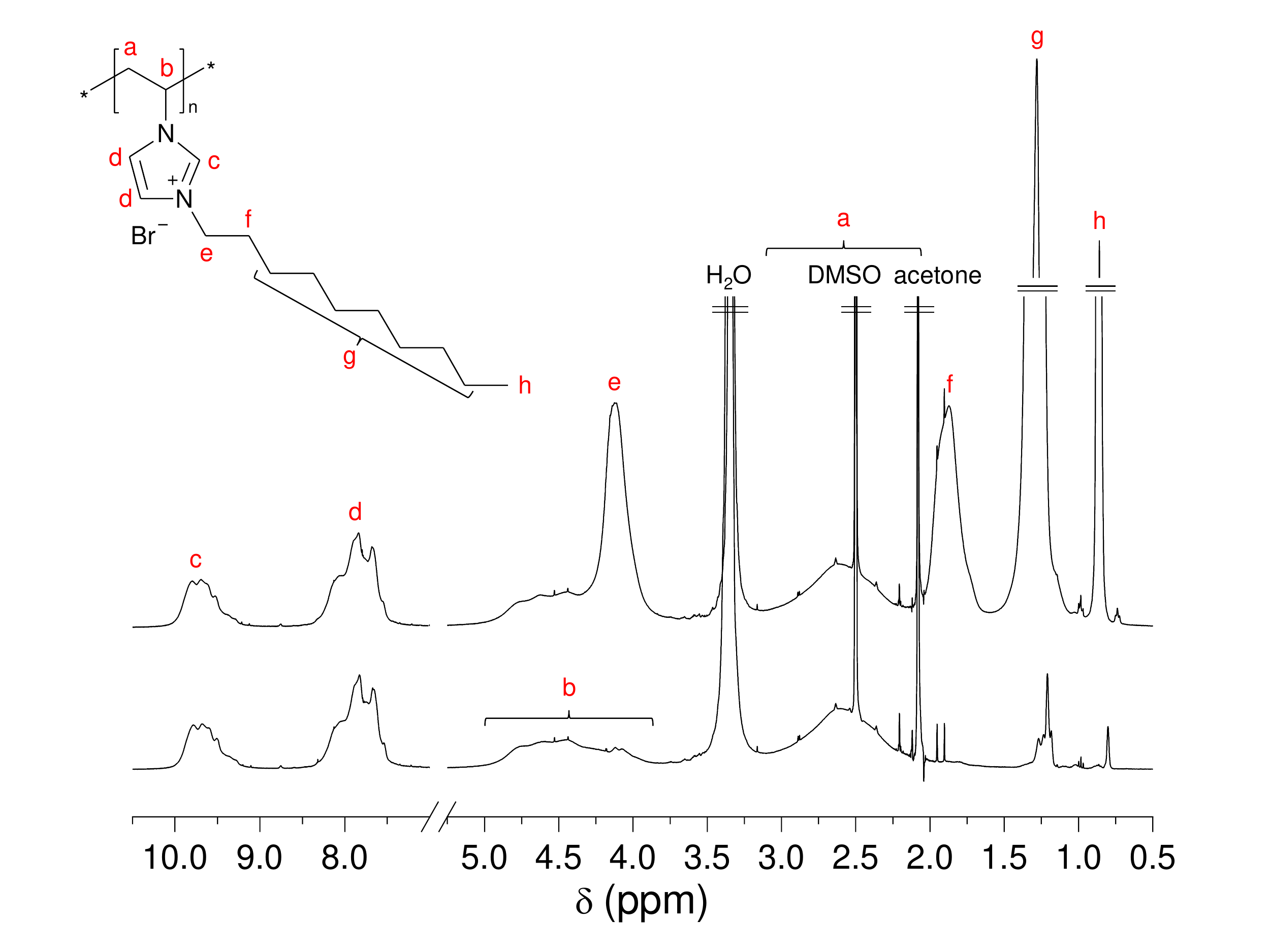}
	\caption{\ce{^{1}H} NMR spectrum (DMSO-\ce{d_6}, 400 MHz) of \textbf{h-PC$_{10}$Br} (top) and \textbf{d-PC$_{10}$Br} (bottom).}
	\label{fig:nmr_BrH10_BrD10}
\end{figure}
\newpage
\subsection{General Procedure for Ion Methathesis}

\textbf{Synthesis of h-PC$_1$TFSI.} A solution of \textbf{h-PC$_1$I} (800 mg, 3.39 mmol of imidazolium iodide groups) and LiTFSI (2.44 g, 8.50 mmol) in 20 mL of a 1:1 (v/v) mixture of acetonitrile and methanol was heated at 40 °C for 24h. The resulting mixture was concentrated under reduced pressure and precipitated in 40 mL of cold water. The crude product was redissolved in acetonitrile and precipitated in cold water to afford after filtration and thorough drying under vacuum \textbf{h-PC$_1$TFSI} as a light yellow solid (1.15 g, 87.1 \%). \ce{^1H} NMR (DMSO-\ce{d_6}, 400 MHz): $\delta$ 9.20–8.26 (1H, br, \ce{H_c}), 8.02–6.67 (2H, br, \ce{H_d}), 4.61–3.49 (4H, br, \ce{H_b}, \ce{H_e}), 2.84–1.61 (2H, br, \ce{H_a}). \ce{^19F} NMR (DMSO-\ce{d_6} with 0.05 \% v/v \ce{CFCl_3}, 376.5 MHz): $\delta$ –78.35 (6F, s, \ce{(CF_3SO_2)_2N}).

\textbf{Synthesis of d-PC$_1$TFSI.} The general procedure for the ion metathesis was applied to a mixture of \textbf{d-PC$_1$I} (595 mg, 2.49 mmol of imidazolium iodide groups) and LiTFSI (1.79 g, 6.23 mmol) to yield \textbf{d-PC$_1$TFSI} as a white solid (870 mg, 89.2 \%). \ce{^1H} NMR (DMSO-\ce{d_6}, 400 MHz): $\delta$ 9.17–8.23 (1H, br, \ce{H_c}), 8.02–6.64 (2H, br, \ce{H_d}), 4.51–3.46 (1H, br, \ce{H_b}), 2.80–1.62 (2H, br, \ce{H_a}). \ce{^19F} NMR (DMSO-\ce{d_6} with 0.05 \% v/v \ce{CFCl_3}, 376.5 MHz): $\delta$ –78.35 (6F, s, \ce{(CF_3SO_2)_2N}).

\begin{figure}[H]
	\includegraphics[width=0.8\textwidth]{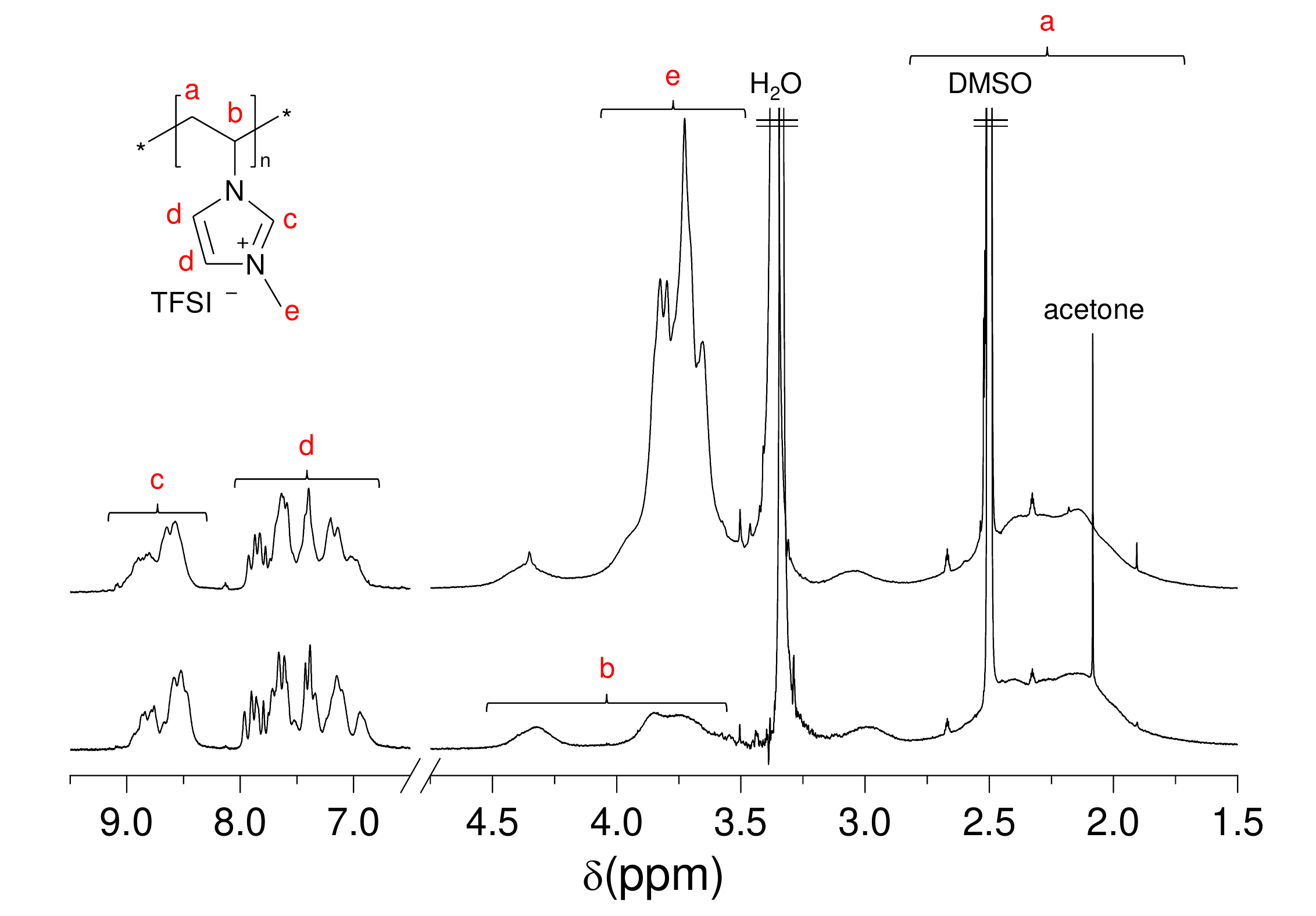}
	\caption{\ce{^{1}H} NMR spectrum (DMSO-\ce{d_6}, 400 MHz) of \textbf{h-PC$_1$TFSI} (top) and \textbf{d-PC$_1$TFSI} (bottom).}
	\label{fig:nmr_TH1_TD1}
\end{figure}
\newpage
\textbf{Synthesis of h-PC$_2$TFSI.} The general procedure for the ion metathesis was applied to a mixture of \textbf{h-PC$_2$I} (615 mg, 2.46 mmol of imidazolium iodide groups) and LiTFSI (1.77 g, 6.17 mmol) to yield \textbf{h-PC$_2$TFSI} as a light yellow solid (805 mg, 81.3 \%). \ce{^1H} NMR (DMSO-\ce{d_6}, 400 MHz): $\delta$ 9.29–8.33 (1H, br, \ce{H_c}), 8.22–6.45 (2H, br, \ce{H_d}), 4.67–3.47 (3H, br, \ce{H_b}, \ce{H_e}), 3.03–1.73 (2H, br, \ce{H_a}), 1.73–1.10 (3H, br, \ce{H_f}). \ce{^19F} NMR (DMSO-\ce{d_6} with 0.05 \% v/v \ce{CFCl_3}, 376.5 MHz): $\delta$ –78.35 (6F, s, \ce{(CF_3SO_2)_2N}).

\textbf{Synthesis of d-PC$_2$TFSI.} The general procedure for the ion metathesis was applied to a mixture of \textbf{d-PC$_2$I} (711 mg, 2.79 mmol of imidazolium iodide groups) and LiTFSI (2.00 g, 6.97 mmol) to yield \textbf{d-PC$_2$TFSI} as a white solid (957 mg, 83.9 \%).\ce{^1H} NMR (DMSO-\ce{d_6}, 400 MHz): $\delta$ 9.29–8.26 (1H, br, \ce{H_c}), 8.18–6.31 (2H, br, \ce{H_d}), 4.64–3.52 (1H, br, \ce{H_b}), 2.99–1.57 (2H, br, \ce{H_a}). \ce{^19F} NMR (DMSO-\ce{d_6} with 0.05 \% v/v \ce{CFCl_3}, 376.5 MHz): $\delta$ –78.35 (6F, s, \ce{(CF_3SO_2)_2N}).

\begin{figure}[H]
	\includegraphics[width=0.8\textwidth]{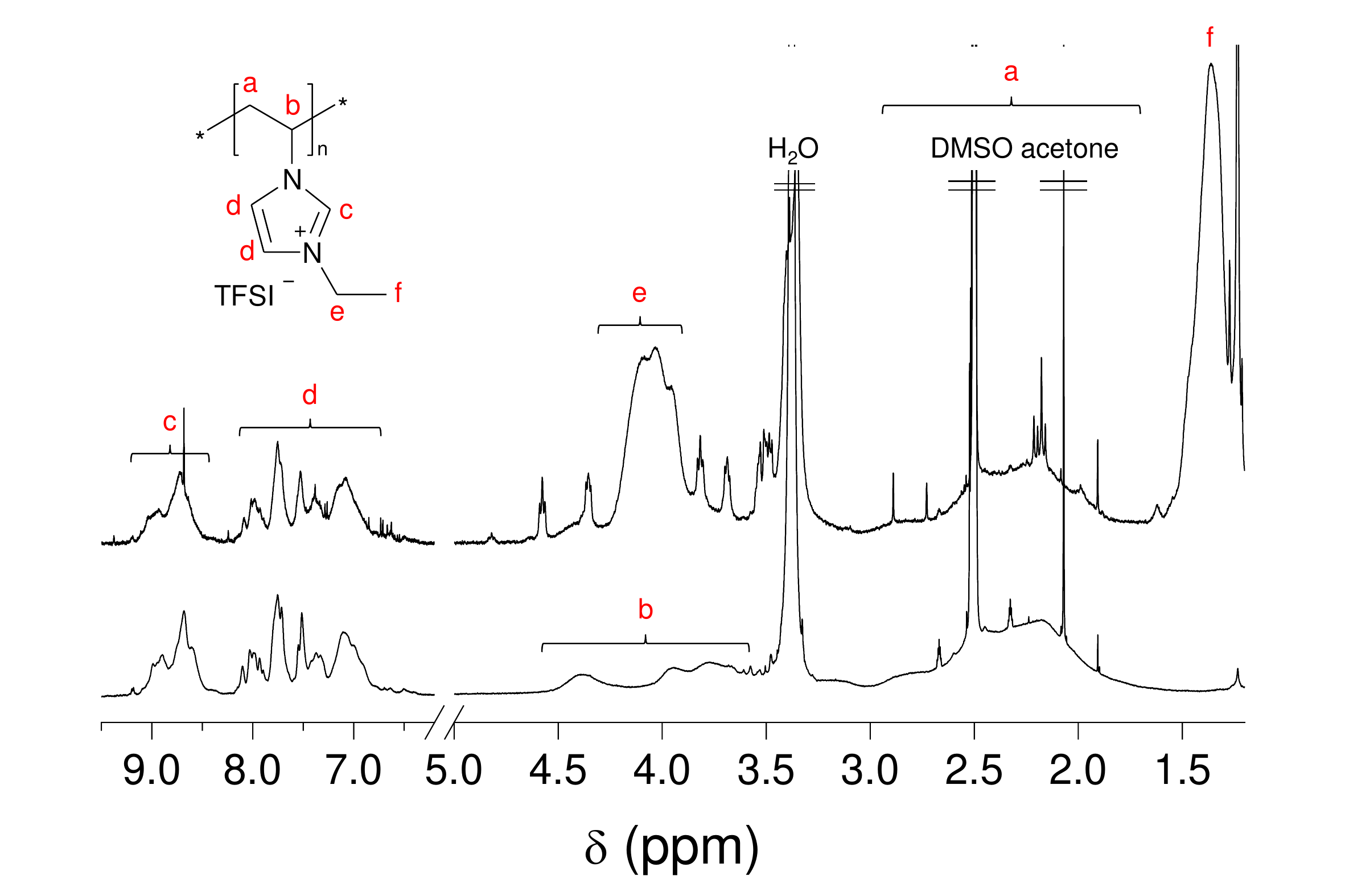}
	\caption{\ce{^{1}H} NMR spectrum (DMSO-\ce{d_6}, 400 MHz) of \textbf{h-PC$_2$TFSI} (top) and \textbf{d-PC$_2$TFSI} (bottom).}
	\label{fig:nmr_TH2_TD2}
\end{figure}
\newpage
\textbf{Synthesis of h-PC$_4$TFSI.} The general procedure for the ion metathesis was applied to a mixture of \textbf{h-PC$_4$I} (968 mg, 3.48 mmol of imidazolium iodide groups) and LiTFSI (2.50 g, 8.71 mmol) to yield \textbf{h-PC$_4$TFSI} as a light yellow solid (1.26 g, 83.9 \%). \ce{^1H} NMR (DMSO-\ce{d_6}, 400 MHz): $\delta$ 9.32-8.31 (1H, br, \ce{H_c}), 8.20–6.34 (2H, br, \ce{H_d}), 4.84–3.56 (3H, br, \ce{H_b}, \ce{H_e}), 3.13–1.92 (2H, br, \ce{H_a}), 1.92–1.48 (2H, br, \ce{H_f}), 1.53–1.48 (2H, br, \ce{H_g}), 1.16–0.72 (3H, br, \ce{H_h}). \ce{^19F} NMR (DMSO-\ce{d_6} with 0.05 \% v/v \ce{CFCl_3}, 376.5 MHz): $\delta$ –78.35 (6F, s, \ce{(CF_3SO_2)_2N}).

\textbf{Synthesis of d-PC$_4$TFSI.} The general procedure for the ion metathesis was applied to a mixture of \textbf{d-PC$_4$I} (970 mg, 3.38 mmol of imidazolium iodide groups) and LiTFSI (2.43 g, 8.46 mmol) to yield \textbf{d-PC$_4$TFSI} as a brown solid (1.35 g, 90.8 \%). \ce{^1H} NMR (DMSO-\ce{d_6}, 400 MHz): $\delta$ 9.30–8.36 (1H, br, \ce{H_c}), 8.17–6.38 (2H, br, \ce{H_d}), 4.74–3.57 (1H, br, \ce{H_b}), 3.11–1.55 (2H, br, \ce{H_a}). \ce{^19F} NMR (DMSO-\ce{d_6} with 0.05 \% v/v \ce{CFCl_3}, 376.5 MHz): $\delta$ –78.35 (6F, s, \ce{(CF_3SO_2)_2N}).

\begin{figure}[H]
	\includegraphics[width=0.8\textwidth]{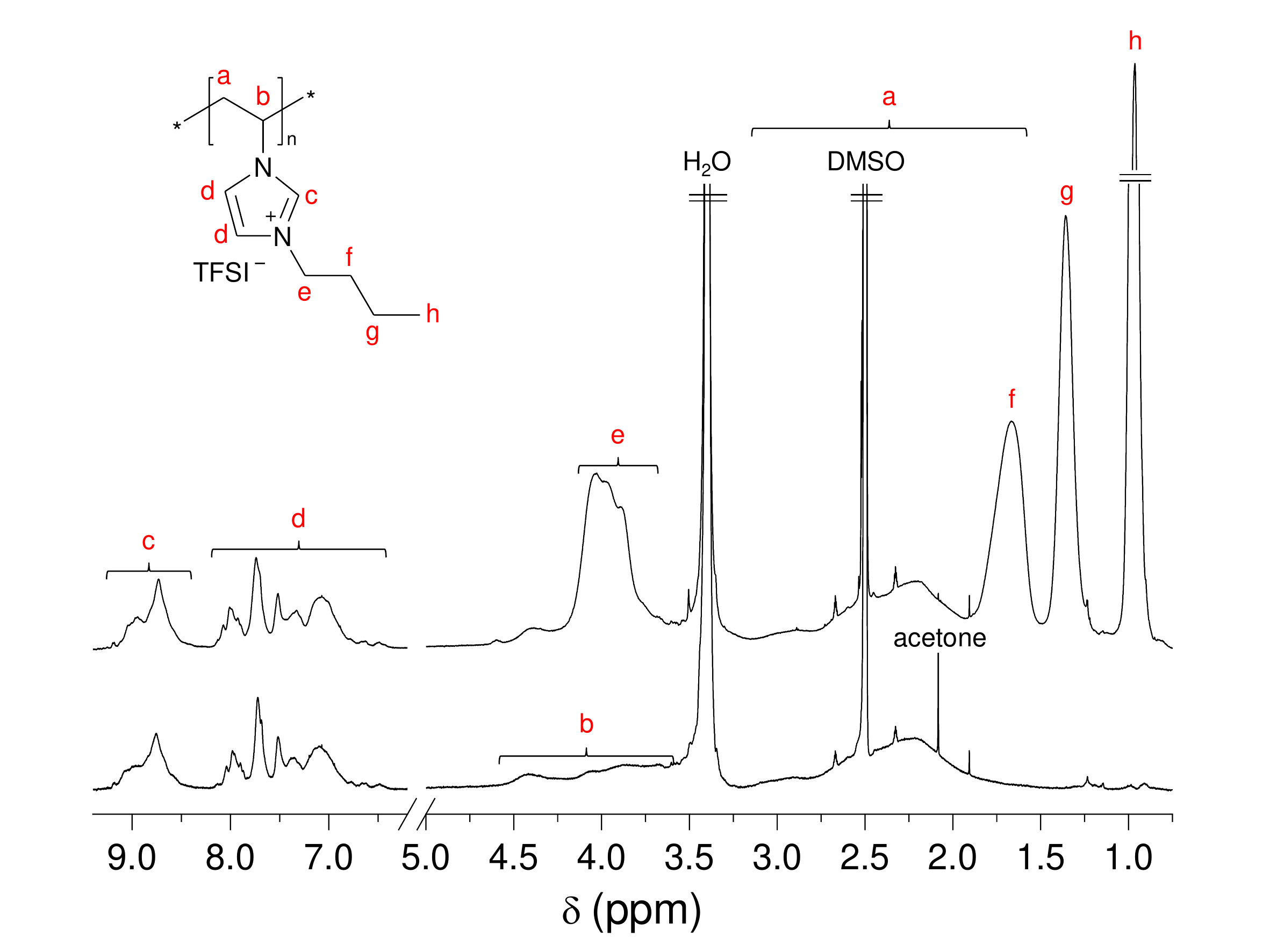}
	\caption{\ce{^{1}H} NMR spectrum (DMSO-\ce{d_6}, 400 MHz) of \textbf{h-PC$_4$TFSI} (top) and \textbf{d-PC$_4$TFSI} (bottom).}
	\label{fig:nmr_TH4_TD4}
\end{figure}
\newpage
\textbf{Synthesis of h-PC$_{10}$TFSI.} The general procedure for the ion metathesis was applied to a mixture of \textbf{h-PC$_{10}$Br} (682 mg, 2.16 mmol of imidazolium bromide groups) and LiTFSI (1.55 g, 5.40 mmol) to yield \textbf{h-PC$_{10}$TFSI} as a light yellow solid (982 mg, 97.2 \%). \ce{^1H} NMR (DMSO-\ce{d_6}, 400 MHz): $\delta$ 9.45–8.33 (1H, br, \ce{H_c}), 8.20–6.60 (2H, br, \ce{H_d}), 4.82–3.56 (3H, br, \ce{H_b}, \ce{H_e}), 3.00–1.96 (2H, br, \ce{H_a}), 2.96–1.52 (2H, br, \ce{H_f}), 1.52–1.04 (14H, br, \ce{H_g}), 1.04–0.66 (3H, br, \ce{H_h}). \ce{^19F} NMR (DMSO-\ce{d_6} with 0.05 \% v/v \ce{CFCl_3}, 376.5 MHz): $\delta$ –78.38 (6F, s, \ce{(CF_3SO_2)_2N}).

\textbf{Synthesis of d-PC$_{10}$TFSI.} The general procedure for the ion metathesis was applied to a mixture of \textbf{d-PC$_{10}$Br} (580 mg, 1.72 mmol of imidazolium bromide groups) and LiTFSI (1.24 g, 4.32 mmol) to yield \textbf{d-PC$_{10}$TFSI} as a light yellow solid (791 mg, 94.2 \%). \ce{^1H} NMR (DMSO-\ce{d_6}, 400 MHz): $\delta$ 9.40–8.38 (1H, br, \ce{H_c}), 8.16–6.73 (2H, br, \ce{H_d}), 4.81–3.56 (1H, br, \ce{H_b}), 3.08–1.52 (2H, br, \ce{H_a}). \ce{^19F} NMR (DMSO-\ce{d_6} with 0.05 \% v/v \ce{CFCl_3}, 376.5 MHz): $\delta$ –78.38 (6F, s, \ce{(CF_3SO_2)_2N}).

\begin{figure}[h!]
	\includegraphics[width=0.8\textwidth]{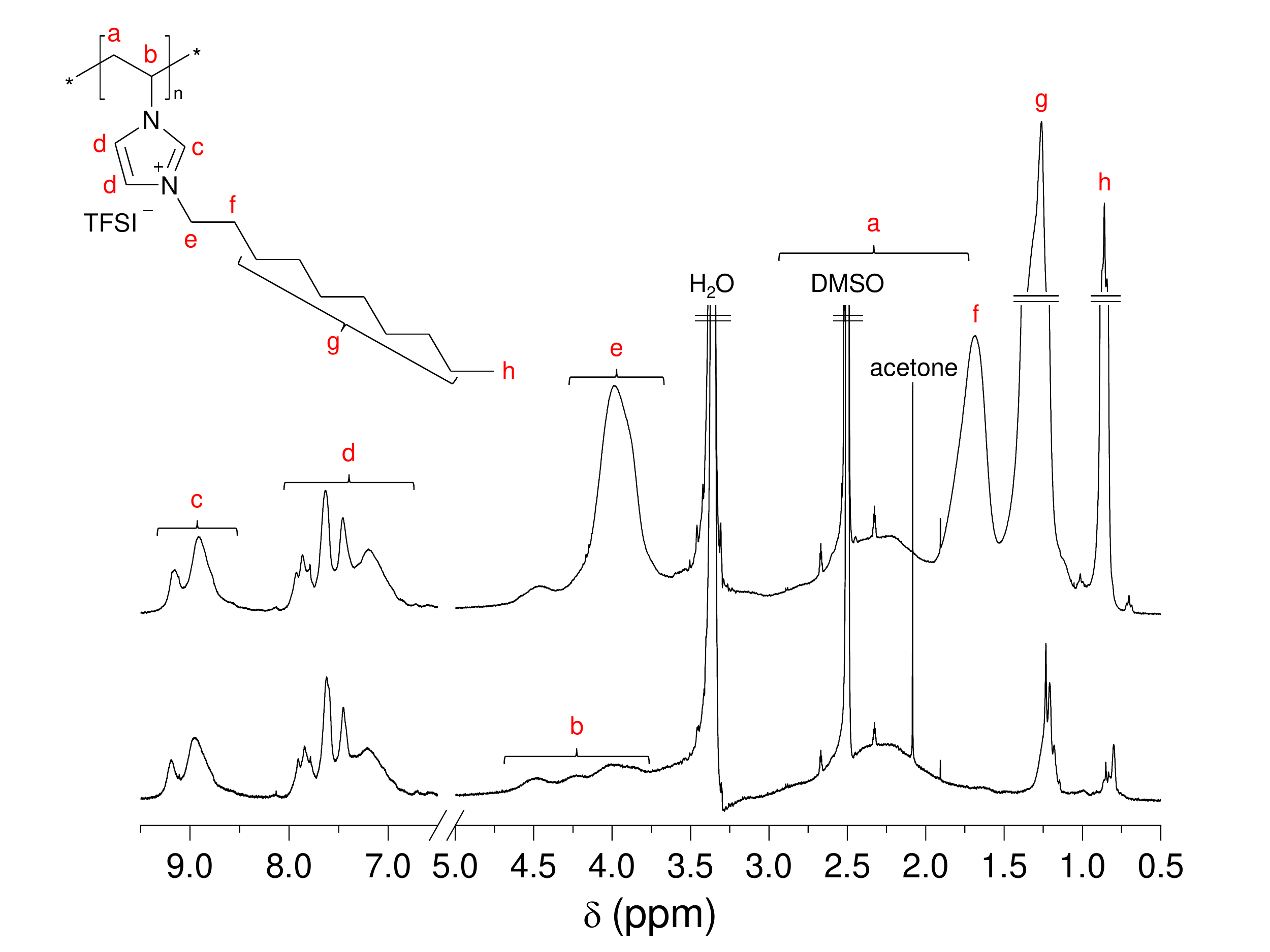}
	\caption{\ce{^{1}H} NMR spectrum (DMSO-\ce{d_6}, 400 MHz) of \textbf{h-PC$_{10}$TFSI} (top) and \textbf{d-PC$_{10}$TFSI} (bottom).}
	\label{fig:nmr_TH10_TD10}
\end{figure}
\newpage
\begin{figure}[H]
	\includegraphics[width=0.8\textwidth]{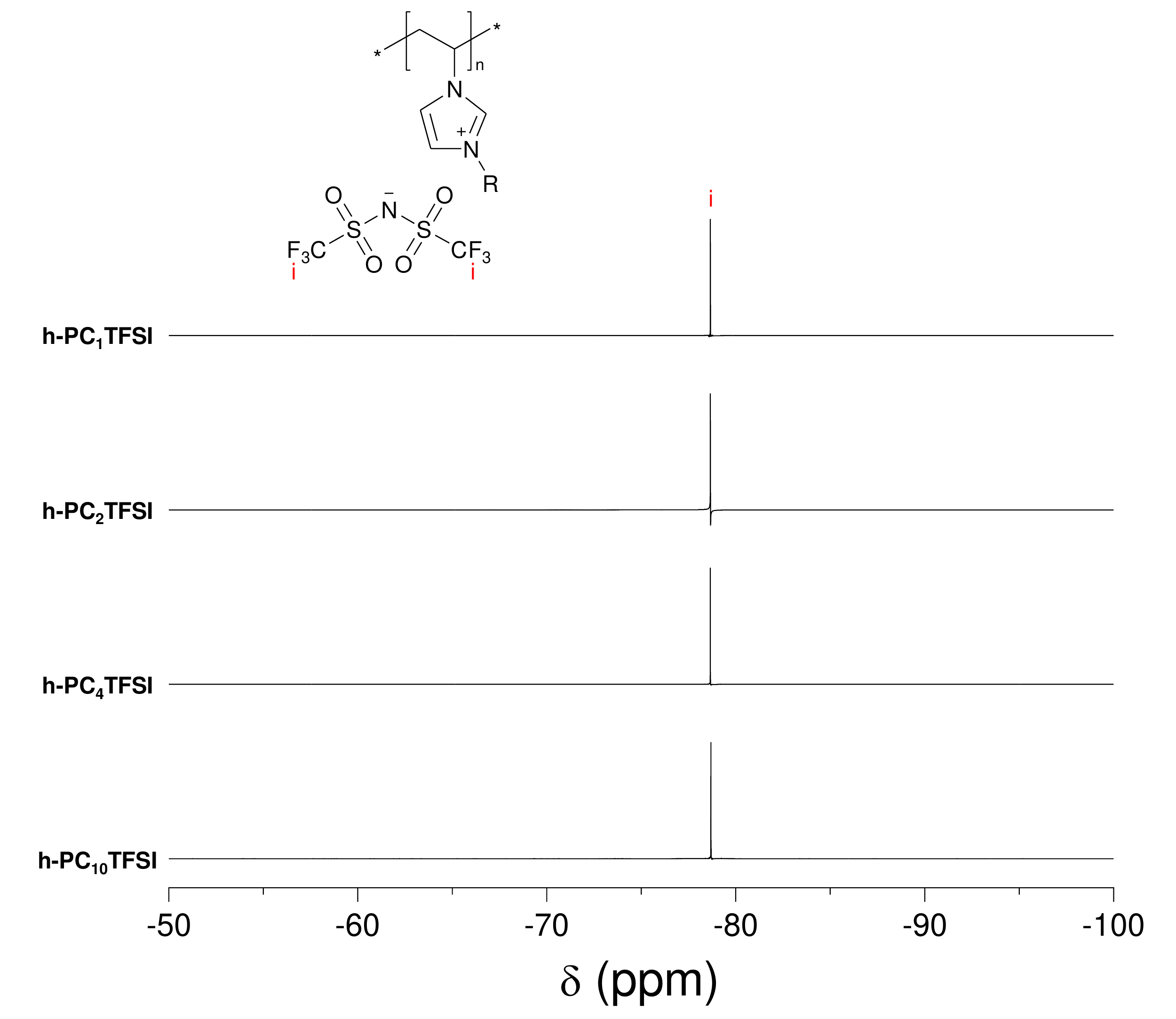}
	\caption{\ce{^{19}F} NMR spectrum (DMSO-\ce{d_6} with 0.05 \% v:v \ce{CFCl_3}, 400 MHz) of \textbf{h-PC$_1$TFSI},  \textbf{h-PC$_2$TFSI}, \textbf{h-PC$_4$TFSI},  \textbf{h-PC$_{10}$TFSI}.}
	\label{fig:F_nmr_TH}
\end{figure}

\begin{figure}[H]
	\includegraphics[width=0.8\textwidth]{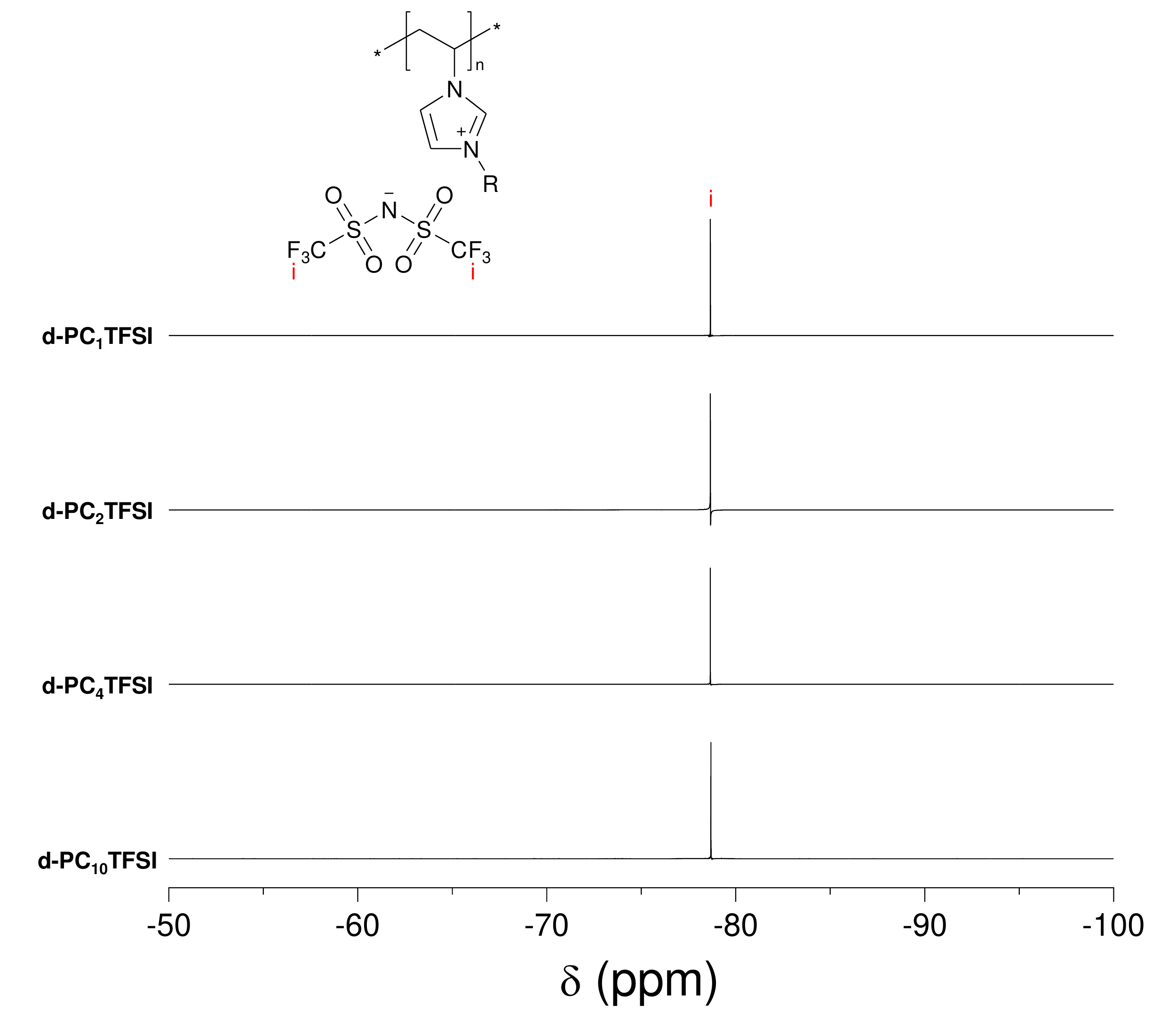}
	\caption{\ce{^{19}F} NMR spectrum (DMSO-\ce{d_6} with 0.05 \% v:v \ce{CFCl_3}, 400 MHz) of \textbf{d-PC$_1$TFSI},  \textbf{d-PC$_2$TFSI}, \textbf{d-PC$_4$TFSI},  \textbf{d-PC$_{10}$VI-TFSI}.}
	\label{fig:F_nmr_TD}
\end{figure}

\begin{figure}[H]
	\includegraphics[width=\textwidth]{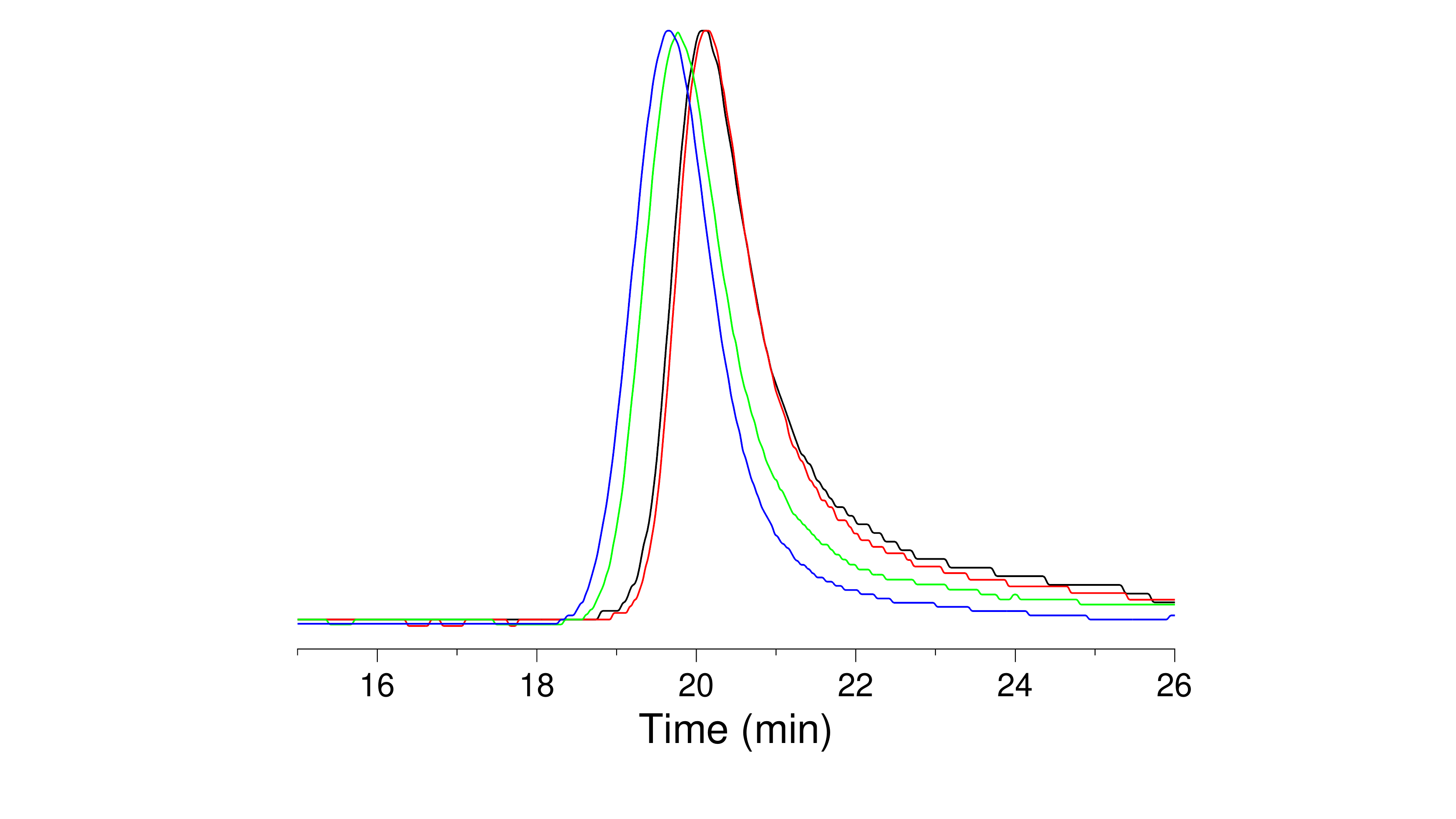}
	\caption{Size exclusion chromatography traces of \textbf{h-PC$_1$TFSI} (black solid line), \textbf{h-PC$_2$TFSI} (red solid line), \textbf{h-PC$_4$TFSI} (green solid line) and \textbf{h-PC$_{10}$TFSI} (blue solid line) in 0.01 M LiTFSI in DMF.}
	\label{fig:SEC}
\end{figure}

\bibliographystyle{achemso}